\documentclass[preprint,12pt]{elsarticle}

\usepackage{amsmath,amsfonts,amssymb,bm}
\usepackage{xcolor}

\usepackage{mathtools}
\usepackage{algorithm,algpseudocode}
\usepackage{tikz,pgfplotstable}
\usepackage{subcaption}
\usepackage{booktabs}
\usepackage{siunitx}
\usepackage[normalem]{ulem}
\usepackage{hyperref}

\colorlet{inlinkcolor}{green!50!black}
\colorlet{exlinkcolor}{red!50!black}
\hypersetup{
  colorlinks = true,
  allcolors = inlinkcolor,
  urlcolor = exlinkcolor,
}

\DeclarePairedDelimiterX{\infdivx}[2]{(}{)}{%
  #1\;\delimsize\|\;#2%
}
\newcommand{\DKL}{D_{\mathrm{KL}}\infdivx}

\DeclareMathOperator*{\argmin}{arg\,min}
\DeclareMathOperator*{\argmax}{arg\,max}
\DeclareMathOperator{\tr}{tr}

\pgfplotstableread{cholesky-factor-pattern/cholesky-factor-pattern-full.dat}\cfpfull
\pgfplotstableread{cholesky-factor-pattern/cholesky-factor-pattern-diag.dat}\cfpdiag
\pgfplotstableread{cholesky-factor-pattern/cholesky-factor-pattern-chev.dat}\cfpchev

\begin{document}

\begin{frontmatter}

  \title{Approximate Bayesian Model Inversion for PDEs with Heterogeneous and State-Dependent Coefficients}

  \author[PNNL]{D.A.~Barajas-Solano\corref{cora}}%
  \ead{David.Barajas-Solano@pnnl.gov}%

  \author[PNNL]{A.M.~Tartakovsky}%
  \ead{Alexandre.Tartakovsky@pnnl.gov}%

  \cortext[cora]{Corresponding author}%
  \address[PNNL]{Pacific Northwest National Laboratory, Richland, WA 99354}

  \begin{abstract}
    We present two approximate Bayesian inference methods for parameter estimation in partial differential equation (PDE) models with space-dependent and state-dependent parameters.
    We demonstrate that these methods provide accurate and cost-effective alternatives to Markov Chain Monte Carlo simulation.
    We assume a parameterized Gaussian prior on the unknown functions, and approximate the posterior density by a parameterized multivariate Gaussian density.
    The parameters of the prior and posterior are estimated from sparse observations of the PDE model's states and the unknown functions themselves by maximizing the evidence lower bound (ELBO), a lower bound on the log marginal likelihood of the observations.
    The first method, Laplace-EM, employs the expectation maximization algorithm to maximize the ELBO, with a Laplace approximation of the posterior on the E-step, and minimization of a Kullback-Leibler divergence on the M-step.
    The second method, DSVI-EB, employs the doubly stochastic variational inference (DSVI) algorithm, in which the ELBO is maximized via gradient-based stochastic optimization, with nosiy gradients computed via simple Monte Carlo sampling and Gaussian backpropagation.
    We apply these methods to identifying diffusion coefficients in linear and nonlinear diffusion equations, %
    and we find that both methods provide accurate estimates of posterior densities and the hyperparameters of Gaussian priors.
    While the Laplace-EM method is more accurate, it requires computing Hessians of the physics model.
    The DSVI-EB method is found to be less accurate but only requires gradients of the physics model.
  \end{abstract}

  \begin{keyword}
    approximate Bayesian inference, model inversion, variational inference, empirical Bayes
  \end{keyword}
    
\end{frontmatter}

\section{Introduction}
\label{sec:intro}

Partial differential equation (PDE) models of many physical systems involve space-dependent parameters and constitutive relationships that are usually only partially observed.
Model inversion aims to estimate these unknown functions of space and the system's state from sparse measurements of the state, associated quantities of interest, and the unknown functions themselves.
Bayesian inference provides a probabilistic framework for model inversion~\cite{stuart_2010}, in which data is assimilated by computing the posterior density of the parameters in terms of the likelihood of the observations given the PDE model and the prior density of the parameters codifying modeling assumptions.
Unlike deterministic parameter estimation methods~\cite{hanke-1997-regularizing,barajassolano-2014-linear}, the Bayesian framework provides a probabilistic characterization of the estimated parameter that can be employed for quantifying uncertainty and evaluating modeling assumptions.
For linear problems with Gaussian likelihoods and Gaussian priors, Bayesian inference can be done exactly (known in the context of state estimation for dynamical systems as the Kalman filter~\cite{evensen-2006-data}).
Unfortunately, physics models define nonlinear maps between the state and the model parameters, preventing carrying out exact inference even in the case of Gaussian likelihoods and Gaussian priors.
The Markov Chain Monte Carlo (MCMC) method is robust for general nonlinear problems but is computationally expensive~\cite{salimans-2015-markov}.
Despite recent advances in Hamiltonian Monte Carlo and ensemble and parallel MCMC~\cite{goodman-2010-ensemble,neiswanger-2013-asymptotically,hoffman-2014-nuts}, the number of forward simulations and likelihood evaluations required by MCMC sampling poses a challenge for model inversion of PDE models with high-dimensional parameters.
Here, we propose two cost-effective alternatives to MCMC for estimating unknown parameters and constitutive relationships in PDE models.

Gaussian process (GP) regression, known as kriging in spatial geophysics, is commonly used to construct probabilistic models of heterogeneous parameters; therefore, GPs serve as a reasonable choice of prior for unknown parameters.
In the context of Bayesian inference with GP priors, GP regression is equivalent to exact Bayesian inference for assimilating direct measurements of unknown parameters.
Similarly, the marginal likelihood of parameter measurements can be computed in closed form, therefore allowing for model selection to be carried out by empirical Bayesian inference, also known as type-II maximum likelihood estimation~\cite{rasmussen-2005-gaussian}.

Assimilating measurements of the state of PDE models is on the other hand less straightforward.
Recently, a framework has been proposed to combine GP priors on the state and a discretization of the governing PDEs to assimilate state observations~\cite{raissi-2018-numerical,raissi-2017-machine}.
In this framework, state estimation and type-II maximum likelihood can be carried out in closed form when the governing equations are linear on the state; for the nonlinear case, inference is carried out approximately via linearization of the governing equations.

Parameter estimation for PDE models presents another layer of challenge as governing equations commonly induce nonlinear relations between parameters and states. 
A common example is the Laplace equation with space-dependent unknown diffusion coefficient, which is linear on the state, but induces a nonlinear relation between the state and the diffusion coefficient.
For the general case of parameter estimation with nonlinearity introduced by the physics model, approximate Bayesian inference methods are necessary.
The standard approximate inference tool is MCMC sampling of the Bayesian posterior.
Given unbounded computational resources, MCMC will provide arbitrarily accurate results, but in practice MCMC often requires an intractable amount of forward simulations of the physics model.
Algorithms such as Hamiltonian Monte Carlo (HMC) and the Metropolis-adjusted Langevin algorithm (MALA) employ first-order information in the form sensitivities of the physics model to improve the mixing and convergence of the Markov chain random walk, but nevertheless the total number of forward and sensitivity simulations remains a challenge.
As an alternative, approaches such as the Laplace approximation and variational inference aim to approximate the exact posterior with an analytical, parameterized density.

In this manuscript we propose employing approximate Bayesian inference with GP priors to approximate the posterior of PDE parameters and to estimate the hyperparameters of their GP prior.
We propose two optimization-based methods: The first, Laplace-EM, is based on the Laplace approximation~\cite{rasmussen-2005-gaussian,bishop-2006-pattern,lawrence-2007-modelling} and the expectation maximization (EM) algorithm~\cite{neal-1998-view,bishop-2006-pattern}.
The second, doubly stochastic variational inference for empirical Bayes inference (DSVI-EB) is based on the DSVI algorithm~\cite{ranganath-2014-black,titsias-2014-doubly}.
The proposed methods employ first and second-order information, i.e., gradient and Hessian of physics models, evaluated via the discrete adjoint method.
Both presented methods enjoy advantageous computational properties over MCMC and other approximate Bayesian inference algorithms such as expectation propagation~\cite{minka-2001-expectation} and the Laplace approximation-based method of~\cite{lawrence-2007-modelling} 
that renders each of them attractive for model inversion depending on the nature of the inversion problem.
In particular, the Laplace-EM method is accurate for approximating the unimodal posteriors of the numerical examples of this manuscript, but requires computing Hessians.
On the other hand, DSVI-EB is less accurate but only requires computing gradients, and can be trivially parallelized.
We note that Gaussian mixtures can be employed in variational inference to approximate multimodal posteriors~\cite{tsilifis-2016-computationally}, but in the present work we limit our focus to unimodal posteriors.
  Furthermore, both methods are applicable to non-factorizing likelihoods, do not require computing moments of the likelihood, and do not require third- or higher order derivatives of the physics model.
Finally, variational inference and the Laplace approximation have been employed for model inversion~\cite{jin-2010-hierarchical,franck-2016-sparse,guha-2015-variational,yang-2017-bayesian,lawrence-2007-modelling}, but to the best of our knowledge, have not been used in the context of the empirical Bayesian framework to estimate GP prior hyperparameters, with the exception of the work of~\cite{lawrence-2007-modelling}.
That work, based on the Laplace approximation, requires computing third-order derivatives of the physics model, which may be costly to compute, whereas the presented methods do not require third-order derivatives.

The manuscript is structured as follows: In Section~\ref{sec:problem} we formulate the empirical Bayesian inference problem for physics models and GP priors.
In Section~\ref{sec:ai-em} we propose the approximate Bayesian inference and summarize the expectation maximization (EM) algorithm.
The Laplace-EM algorithm is introduced in Section~\ref{sec:laplace-em}, and the DSVI-EB algorithm is described in~Section~\ref{sec:dsvi}.
The computational complexity of the algorithms is discussed in~Section~\ref{sec:cost}.
The application of the proposed methods are presented in Section~\ref{sec:numexp}.
Finally, conclusions are given in~Section~\ref{sec:conclusions}.

\section{Problem formulation}
\label{sec:problem}

We consider physical systems modeled by a stationary PDEs over the simulation domain $\Omega \subset \mathbb{R}^d$, $d \in [1, 3]$.
We denote by $u \colon \Omega \to U \subset \mathbb{R}$ the system's state, and by $y \colon \Omega \times U \to \mathbb{R}$ the system's parameter, an unknown scalar function of space and the system's state.
Our goal is to estimate the unknown function $y(x, u)$ from sparse, noisy measurements of $u(\mathbf{x})$ and $y(\mathbf{x}, u)$.
The PDE and boundary conditions are discretized for numerical computations, resulting in the set of $M$ algebraic equations $\mathbf{L}(\mathbf{u}, \mathbf{y}) = 0$, where $\mathbf{u} \in \mathbb{R}^M$ denotes the vector of $M$ state degrees of freedom, and $\mathbf{y} \in \mathbb{R}^N$ denotes the discretized parameter vector, corresponding to the value of $y(\mathbf{x}, u)$ at the $N$ discrete locations $\{ \xi_i \in \Omega \times U \}^N_{i = 1}$.
Furthermore, we denote by $\bm{\Xi}$ the matrix of coordinates $\bm{\Xi} \equiv (\xi_1, \dots, \xi_N)$.

We assume that the sparse observations of the discrete state and parameters, $\mathbf{u}_{\mathrm{s}}$ and $\mathbf{y}_{\mathrm{s}}$, respectively, are collected with iid normal observation errors, that is,
\begin{gather}
  \label{eq:uobs}
  \mathbf{u}_{\mathrm{s}} = \mathbf{H}_u \mathbf{u} + \bm{\epsilon}_u, \quad \bm{\epsilon}_{us} \sim \mathcal{N}(0, \sigma^2_{us} \mathbf{I}_{M_s}),\\
  \label{eq:yobs}
  \mathbf{y}_{\mathrm{s}} = \mathbf{H}_y \mathbf{y} + \bm{\epsilon}_y, \quad \bm{\epsilon}_{ys} \sim \mathcal{N}(0, \sigma^2_{ys} \mathbf{I}_{N_s}),
\end{gather}
where $\mathbf{u}_{\mathrm{s}} \in \mathbb{R}^{M_s}$, $M_s \ll M$ are the state observations, $\mathbf{y}_{\mathrm{s}} \in \mathbb{R}^{N_s}$, $N_s \ll N$ are the parameter observations, $\mathbf{H}_u \in \mathbb{R}^{M_s \times M}$ is the state observation operator, $\mathbf{H}_y \in \mathbb{R}^{N_s \times N}$ is the parameter observation operator, and $\bm{\epsilon}_{us}$ and $\bm{\epsilon}_{ys}$ are observation errors satisfying $\mathbb{E}[ \bm{\epsilon}_{us} \bm{\epsilon}^{\top}_{ys}] = 0$.
Then, the likelihood of the observations $\mathcal{D}_s \equiv \{ \mathbf{u}_s, \mathbf{y}_s \}$ given $\mathbf{y}$ is defined as
\begin{equation}
  \label{eq:likelihood}
  \log p(\mathcal{D}_s \mid \mathbf{y}) \equiv -\frac{1}{2 \sigma^2_{us}} \| \mathbf{u}_s - \mathbf{H}_u \mathbf{u} \|^2_2 - \frac{1}{2 \sigma^2_{ys}} \| \mathbf{y}_s - \mathbf{H}_y \mathbf{y} \|^2_2 + \text{const.},
\end{equation}
where $\mathbf{u}$ satisfies the physics constraint $\mathbf{L}(\mathbf{u}, \mathbf{y}) = 0$ given $\mathbf{y}$, and the constant is independent of $\mathbf{y}$.

In probabilistic terms, our goal is to estimate the posterior density of $\mathbf{y}$ given the data $\mathcal{D}_s$.
By Bayes' theorem, this posterior is given by
\begin{equation}
  \label{eq:bayes}
  p(\mathbf{y} \mid \mathcal{D}_s, \bm{\theta}) = \frac{p(\mathcal{D}_s \mid \mathbf{y}) p(\mathbf{y} \mid \bm{\theta})}{p(\mathcal{D}_s \mid \bm{\theta})},
\end{equation}
where $p(\mathbf{y} \mid \bm{\theta})$ is the parameterized prior density of $\mathbf{y}$, with hyperparameters $\bm{\theta}$, and $p(\mathcal{D}_s \mid \bm{\theta})$ is the \emph{marginal likelihood} or \emph{evidence} of the data, given by
\begin{equation}
  \label{eq:ml}
  p(\mathcal{D}_s \mid \bm{\theta}) = \int p(\mathcal{D}_s \mid \mathbf{y}) p(\mathbf{y} \mid \bm{\theta}) \, \mathrm{d} \mathbf{y}.
\end{equation}
If one is not interested in the uncertainty in estimating $\mathbf{y}$ given the data, one can compute in lieu of the full posterior (\eqref{eq:bayes}) the \emph{maximum a posteriori} (MAP) point estimate of $\mathbf{y}$, defined as the mode of the posterior, that is,
\begin{equation}
  \label{eq:map}
  \hat{\mathbf{y}} \equiv \argmax_{\mathbf{y}} \log p(\mathbf{y} \mid \mathcal{D}_s, \bm{\theta}) = \argmax_{\mathbf{y}} \log p(\mathbf{y}, \mathcal{D}_s \mid \bm{\theta}),
\end{equation}
where $p(\mathbf{y}, \mathcal{D}_s \mid \bm{\theta}) = p(\mathcal{D}_s \mid \mathbf{y}) p(\mathbf{y} \mid \bm{\theta})$ is the joint density of the data and the parameters given $\bm{\theta}$.
Here we used the fact that the marginal likelihood $p(\mathcal{D}_s \mid \bm{\theta})$ is independent of $\mathbf{y}$.

We employ a zero-mean Gaussian process prior, that is,
\begin{equation}
  \label{eq:GP-prior}
  p(\mathbf{y} \mid \bm{\theta}) = \mathcal{N}(\mathbf{y} \mid 0, \mathbf{C}_p(\bm{\theta}) \equiv C(\bm{\Xi}, \bm{\Xi} \mid \bm{\theta})),
\end{equation}
where $\mathcal{N}(\cdot \mid \bm{\mu}, \bm{\Sigma})$ denotes the multivariate normal density with mean $\bm{\mu}$ and covariance matrix $\bm{\Sigma}$, and $C(\cdot, \cdot \mid \bm{\theta})$ is a parameterized covariance kernel.

The posterior density depends on the prior hyperparameters, which can be chosen based on prior expert knowledge, or estimated from data.
In the empirical Bayes approach, also known as \emph{type-II maximum likelihood} or marginal likelihood estimation, point estimates of the hyperparameters are obtained by maximizing the marginal likelihood with respect to $\bm{\theta}$, i.e., $\hat{\bm{\theta}} \equiv \argmax_{\bm{\theta}} p(\mathcal{D}_s \mid \bm{\theta})$.
In the fully Bayes approach, we instead pose a hyperprior on the hyperparameters, which is updated with data by the Bayes' theorem.
In this work we will pursue the empirical Bayes approach.

Due to the nonlinear map from $\mathbf{y}$ to $\mathbf{u}$ defined by the physics constraint, the Bayesian inference problem of evaluating the posterior and marginal likelihood cannot be done in closed form.
Exact inference therefore requires sampling the posterior via MCMC, which is intractable for sufficiently large $N$ and $M$.
As a consequence, estimating hyperparameters via marginal likelihood estimation is also intractable.
As an alternative to exact inference, in this work we propose various approximate inference algorithms.

\section{Approximate inference and Expectation Maximization}
\label{sec:ai-em}

The goal is to  approximate the exact posterior $p(\mathbf{y} \mid \mathcal{D}_s, \bm{\theta})$ by a density $q(\mathbf{y})$.
The Kullback-Leibler (KL) divergence $\DKL{q(\mathbf{y})}{p(\mathbf{y} \mid \mathcal{D}_s, \bm{\theta})}$ provides a means to rewriting the marginal likelihood, \eqref{eq:ml}, in terms of $q(\mathbf{y})$.
Namely, substituting \eqref{eq:bayes} into the definition of the KL divergence gives
\begin{equation}
  \label{eq:kl-q-p}
  \begin{split}
    \DKL{q(\mathbf{y})}{p(\mathbf{y} \mid \mathcal{D}_s, \bm{\theta})} &= - \int q(\mathbf{y}) \log \frac{p(\mathbf{y} \mid \mathcal{D}_s, \bm{\theta})}{q(\mathbf{y})} \, \mathrm{d} \mathbf{y}\\
    &= - \int q(\mathbf{y}) \log \frac{p(\mathcal{D}_s \mid \mathbf{y}) p(\mathbf{y} \mid \bm{\theta})}{q(\mathbf{y}) p (\mathcal{D}_s \mid \bm{\theta})} \, \mathrm{d} \mathbf{y} \\
    &= - \mathcal{F}\left [ q(\mathbf{y}), \bm{\theta} \right ] + \log p(\mathcal{D}_s \mid \bm{\theta}).
  \end{split}
\end{equation}
where $\mathcal{F} \left [q(\mathbf{y}), \bm{\theta} \right ]$ is given by
\begin{equation}
  \label{eq:elbo}
  \mathcal{F} \left [ q(\mathbf{y}), \bm{\theta} \right ] = \mathbb{E}_{q(\mathbf{y})} \left [ p(\mathcal{D}_s \mid \mathbf{y}) \right ] - \DKL{q(\mathbf{y})}{p(\mathbf{y} \mid \bm{\theta})},
\end{equation}
and $\mathbb{E}_{q(\mathbf{y})} \left [ \cdot \right ] \equiv \int ( \cdot ) q(\mathbf{y}) \, \mathrm{d} \mathbf{y}$ denotes expectation with respect to the density $q(\mathbf{y})$.
Reorganizing~\eqref{eq:kl-q-p} we have the following alternative expression for~\eqref{eq:ml}:
\begin{equation*}
  \log p(\mathcal{D}_s \mid \bm{\theta}) = \mathcal{F} \left [ q(\mathbf{y}), \bm{\theta} \right ] + \DKL{q(\mathbf{y})}{p(\mathbf{y} \mid \mathcal{D}_s, \bm{\theta})}.
\end{equation*}
Given that the KL divergence is always non-negative, we have the inequality $\log p(\mathcal{D}_s \mid \bm{\theta}) \geq \mathcal{F} \left [ q(\mathbf{y}), \bm{\theta} \right ]$;
therefore, the operator $\mathcal{F}$ is often called the \emph{evidence lower bound} (ELBO).
The inequality becomes an equality when $q(\mathbf{y}) = p(\mathbf{y} \mid \mathcal{D}_s, \bm{\theta})$, that is, when the variational density is equal to the exact posterior.
In the empirical Bayes setting, this suggest the strategy of selecting both $q$ and $\bm{\theta}$ by maximizing the ELBO~\cite{neal-1998-view}, i.e.,
\begin{equation}
  \label{eq:elbo-max}
  ( \hat{q}(\mathbf{y}), \hat{\bm{\theta}} ) \equiv \argmax_{\left ( q(\mathbf{y}), \bm{\theta} \right )} \mathcal{F} \left [ q(\mathbf{y}), \bm{\theta} \right ].
\end{equation}
Instead of maximizing over $q(\mathbf{y})$ and $\bm{\theta}$ simultaneously, we can do it iteratively by alternating two maximization steps, resulting in the iterative scheme
\begin{equation}
  \label{eq:elbo-EM}
  \begin{aligned}
    & \text{E-step:} & \hat{q}^{(j + 1)}(\mathbf{y}) \text{ set as } \argmax_{q(\mathbf{y})} \mathcal{F} \left [ q(\mathbf{y}), \bm{\theta}^{(j)} \right ] \\
    & \text{M-step:} & \hat{\bm{\theta}}^{(j + 1)} \text{ set as } \argmax_{\theta} \mathcal{F} \left [q^{(j + 1)}(\mathbf{y}), \bm{\theta} \right ].
  \end{aligned}
\end{equation}
thus recovering the expectation maximization (EM) algorithm~\cite{neal-1998-view}.

It remains to specify how the maximization problems~\eqref{eq:elbo-max} and~\eqref{eq:elbo-EM} will be solved, particularly with respect to how to optimize over the space of possible densities $q(\mathbf{y})$ approximating the true posterior.
The approximate inference algorithms presented in this manuscript are based on two families of approximations of the posterior.
The Laplace-EM algorithm (Section~\ref{sec:laplace-em}) uses a \emph{local} approximation around the MAP for a given $\bm{\theta}$, and optimizes the ELBO using the EM algorithm, \eqref{eq:elbo-EM}.
The DSVI-EB algorithm (Section~\ref{sec:dsvi}) uses a parameterized density $q(\mathbf{y} \mid \bm{\phi})$ with \emph{variational} parameters $\bm{\phi}$ to be selected jointly with $\bm{\theta}$ via stochastic optimization.

\section{Laplace-EM algorithm}
\label{sec:laplace-em}

The Laplace approximation is an approach for approximating unimodal posteriors densities.
It consists of fitting a multivariate Gaussian density around the MAP for a given choice of hyperparameters $\bm{\theta}$.
The $j$th E-step of the EM algorithm, \eqref{eq:elbo-EM}, consists of finding the posterior for a given set of hyperparameters, $\bm{\theta}^{(j)}$.
This suggests we can replace the E-step by a Laplace approximation to the posterior, giving raise to the Laplace-EM algorithm.

We proceed to briefly describe the Laplace approximation.
Expanding up to second order the log posterior (see~\eqref{eq:bayes}) around the MAP (\eqref{eq:map}) yields
\begin{multline*}
  \log p(\mathbf{y} \mid \mathcal{D}_s, \bm{\theta}) = - \log p(\mathcal{D}_s \mid \bm{\theta}) + \log p(\hat{\mathbf{y}}, \mathcal{D}_s \mid \bm{\theta})  \\
  + \frac{1}{2} (\mathbf{y} - \hat{\bm{y}})^{\top} \left [ \left . \nabla \nabla \log p(\mathbf{y}, \mathcal{D}_s \mid \bm{\theta}) \right |_{\mathbf{y} = \hat{\mathbf{y}}} \right ] (\mathbf{y} - \hat{\mathbf{y}}) + \dots,
\end{multline*}
where $\left . \nabla \nabla \log p(\mathbf{y}, \mathcal{D}_s \mid \bm{\theta}) \right |_{\mathbf{y} = \hat{\mathbf{y}}}$ denotes the Hessian of the log joint density $\log p(\mathbf{y}, \mathcal{D}_s \mid \bm{\theta})$ around the MAP.
This quadratic expression suggests approximating the posterior by the multivariate Gaussian density $q(\mathbf{y}) \equiv \mathcal{N}(\mathbf{y} \mid \bm{\mu}_q, \bm{\Sigma}_q)$ with mean $\bm{\mu}_{q}$ given by the MAP and covariance $\bm{\Sigma}_q$ given by the Hessian of the log joint density.
In other words, we have
\begin{equation}
  \label{eq:laplace-mean}
  \begin{split}
    \bm{\mu}_q &\equiv \argmin_{\mathbf{y}} \left [ - \log p(\mathbf{y}, \mathcal{D}_s \mid \bm{\theta}) \right ]\\
    &= \argmin_{\mathbf{y}} \left \{ - \log p(\mathcal{D}_s \mid \mathbf{y}) + \frac{1}{2} \left [ \mathbf{y}^{\top} \mathbf{C}^{-1}_{p}(\bm{\theta}) \mathbf{y} + \log \det \mathbf{C}_p(\bm{\theta}) + N \log 2 \pi \right ] \right \},
  \end{split}
\end{equation}
and
\begin{equation}
  \label{eq:laplace-covar}
  \bm{\Sigma}_q \equiv - \left . \nabla \nabla \log p(\mathbf{y}, \mathcal{D}_s \mid \bm{\theta}) \right |_{\mathbf{y} = \hat{\mathbf{y}}} = \mathbf{H} + \mathbf{C}^{-1}_p(\bm{\theta}),
\end{equation}
where $\mathbf{H} \equiv - \left . \nabla \nabla \log p(\mathcal{D}_s \mid \mathbf{y}) \right |_{\mathbf{y} = \bm{\mu}_q}$ denotes the Hessian of the likelihood around $\bm{\mu}_q$.
Note that $\bm{\Sigma}_q$ and $\bm{\mu}_q$ depend on $\bm{\theta}$ indirectly through the dependence of the MAP on $\bm{\theta}$.

In this work we solve the minimization problem~\eqref{eq:laplace-mean} via gradient-based optimization.
The necessary gradient of $\log p(\mathcal{D}_s \mid \mathbf{y})$ with respect to $\mathbf{y}$ is computed via the discrete adjoint method described in \ref{sec:da}.
The Hessian of $\log p(\mathcal{D}_s \mid \mathbf{y})$ with respect to $\mathbf{y}$, necessary to evaluate \eqref{eq:laplace-covar}, is also computed via the discrete adjoint method.

We propose the Laplace-EM algorithm, where the Laplace approximation provides an approximation to the E-step.
For the M-step, we keep the Laplace approximation fixed and maximize the ELBO with respect to the hyperparameters of the GP prior, $\bm{\theta}$.
From~\eqref{eq:elbo} we see that for fixed $q(\mathbf{y})$, $\bm{\theta}$ appears only through the KL divergence $\DKL{q(\mathbf{y})}{p(\mathbf{y} \mid \bm{\theta})}$; therefore, it suffices to minimize this KL divergence at the M-step.
The Laplace-EM M-step reads
\begin{equation}
  \label{eq:laplace-m-step}
  \bm{\theta}^{(j + 1)} = \argmin_{\bm{\theta}} \DKL{q(\mathbf{y})}{p (\mathbf{y} \mid \bm{\theta}^{(j)} )}.
\end{equation}
For the GP prior, this KL divergence is given in closed form by
\begin{equation}
  \label{eq:laplace-DKL}
  \DKL{q(\mathbf{y})}{p(\mathbf{y} \mid \bm{\theta})} = \frac{1}{2} \left [ \tr \left (\mathbf{C}^{-1}_p \bm{\Sigma}_q \right ) + \hat{\mathbf{y}}^{\top} \mathbf{C}^{-1}_p \hat{\mathbf{y}} \vphantom{\log \frac{\det \mathbf{C}_p}{\bm{\Sigma}_q}} - N + \log \frac{\det \mathbf{C}_p}{\operatorname{\det} \bm{\Sigma}_q} \right ].
\end{equation}
If using a gradient method, the gradient of the KL divergence is given by
\begin{equation}
  \label{eq:laplace-DKL-grad}
 \frac{\partial}{\partial \theta_i} \DKL{q(\mathbf{y})}{p(\mathbf{y} \mid \bm{\theta})} = -\frac{1}{2} \hat{\mathbf{y}}^{\top} \mathbf{C}^{-1}_p \frac{\partial \mathbf{C}_p}{\partial \theta_i} \mathbf{C}^{-1}_p \hat{\mathbf{y}} + \frac{1}{2} \tr \left [ \mathbf{C}^{-1}_p \frac{\partial \mathbf{C}_p}{\partial \theta_i} \left (\mathbf{I}_N - \mathbf{C}^{-1}_p \bm{\Sigma}_q \right ) \right].
\end{equation}

The Laplace-EM algorithm is summarized in Algorithm~\ref{alg:laplace-em}.
In practice, the EM iterations are halted once either a maximum number of iterations are completed, or once the relative change in hyperparameters is below a certain threshold, that is, when
\begin{equation*}
  \max \left \{ \left | \theta^{(j + 1)}_i - \theta^{(j)}_i \right | / \left | \theta^{\mathrm{s}}_i \right | \right \}^{N_{\theta}}_{i = 1} \leq \mathrm{rtol},
\end{equation*}
where $N_{\theta}$ is the number of prior hyperparameters, the $\theta^s_i$, $i \in [1, N_{\theta}]$ are prescribed hyperparameter scales (that provide a sense of the magnitude of the hyperparameters), and $\mathrm{rtol}$ is the prescribed tolerance.

\begin{algorithm}
  \caption{Laplace-EM}
  \label{alg:laplace-em}
  \begin{algorithmic}[0]
    \Require $\bm{\theta}^{(0)}$, $\mathbf{C}_p(\bm{\theta})$, $\log p(\mathcal{D}_s \mid \mathbf{y})$
    \State $j \leftarrow 0$
    \Repeat
    \State Compute $\bm{\mu}_q$ using~\eqref{eq:laplace-mean}
    \State Compute $\bm{\Sigma}_q$ using~\eqref{eq:laplace-covar} \Comment E-step
    \State Solve \eqref{eq:laplace-m-step} for $\bm{\theta}^{(j + 1)}$ \Comment M-step
    \State $j \leftarrow j + 1$
    \Until{Convergence}
  \end{algorithmic}
\end{algorithm}

We note that the Laplace approximation is a commonly used tool for unimodal non-Gaussian inference~\cite{rasmussen-2005-gaussian,bishop-2006-pattern,lawrence-2007-modelling}.
Directly maximizing with respect to $\bm{\theta}$, the Laplace approximation to the marginal likelihood  requires evaluating third-order derivatives of the log-likelihood function~\eqref{eq:likelihood} with respect to $\mathbf{y}$. This is due to the implicit dependence of the MAP on $\bm{\theta}$, which requires evaluating third-order derivatives of the physics constraint.
The use of the expectation maximization algorithm allows us to side-step the need of third-order derivatives.
Other methods for non-Gaussian inference such as expectation propagation~\cite{minka-2001-expectation} require multiple evaluations of the moments of the likelihood function, and are therefore not considered in this work.

\section{Doubly stochastic variational inference}
\label{sec:dsvi}

In variational inference (VI)~\cite{blei-2017-variational}, we restrict our choice of $q(\mathbf{y})$ to a parameterized family $q(\mathbf{y} \mid \bm{\phi})$.
In this context we refer to $q$ as the \emph{variational} density and $\bm{\phi}$ as the \emph{variational} parameters.
Following~\eqref{eq:elbo-max}, we will estimate the variational parameters and the GP prior hyperparameters simultaneously by maximizing the corresponding ELBO, that is,
\begin{equation}
  \label{eq:dsvi-est}
  (\hat{\bm{\phi}}, \hat{\bm{\theta}}) = \argmax_{(\bm{\phi}, \bm{\theta})} \mathcal{F}\left [ q(\mathbf{y} \mid \bm{\phi}), \bm{\theta}) \right ].
\end{equation}

In this section we present our proposed implementation of variational inference for empirical Bayes.
The main challenges of VI are (i) approximating the expectations on the expression for the ELBO, \eqref{eq:elbo}, and (ii) optimizing such approximations.
To address these challenges we employ the \emph{doubly stochastic variational inference} (DSVI) framework~\cite{ranganath-2014-black,titsias-2014-doubly}, in which a noisy simple Monte Carlo estimate of the ELBO (1st source of stochasticity) is minimized via a gradient-based stochastic optimization algorithm (2nd source of stochasticity).
In particular, we employ stochastic gradient ascent with the adaptive step-size sequence proposed by \cite{kucukelbir-2017-automatic}.
The gradients of the ELBO estimate with respect to variational parameters and prior hyperparameters are computed via Gaussian backpropagation~\cite{kingma-2013-auto,titsias-2014-doubly,rezende-2014-stochastic,kucukelbir-2017-automatic}.

\subsection{Gaussian backpropagation}
\label{sec:dsvi-gaussian-backprop}

To maximize the ELBO via gradient-based stochastic optimization, we construct unbiased estimates of the ELBO and its gradients with respect to $\bm{\phi}$ and $\bm{\theta}$.
Computing the gradient $\nabla_{\bm{\phi}} \mathcal{F}$ is not trivial as it involves expectations over $q$, which depends on $\bm{\phi}$.

We restrict ourselves to the multivariate Gaussian variational family $q(\mathbf{y} \mid \bm{\phi}) = \mathcal{N}(\mathbf{y} \mid \bm{\mu}_q, \bm{\Sigma}_q)$, with variational mean $\bm{\mu}_q \in \mathbb{R}^N$ and covariance $\bm{\Sigma}_q = \mathbf{R}_q \mathbf{R}_q^{\top} \in \mathbb{R}^{N \times N}$, where $\mathbf{R}_q$ is a lower triangular factor matrix.
For this choice we have $\bm{\phi} = \{\bm{\mu}_q, \mathbf{R}_q \}$.
Similar to the Laplace approximation, this choice is justified for unimodal posteriors.
We then introduce the change of variables $\mathbf{y} = \bm{\mu}_q + \mathbf{R}_q \mathbf{z}$, with $\mathbf{z} \sim \mathcal{N}(0, \mathbf{I}_N)$.
Substituting this change of variables into~\eqref{eq:elbo}, we can rewrite the involved expectations in terms of expectations over $\mathcal{N}(\mathbf{z} \mid 0, \mathbf{I}_N)$, resulting in
\begin{multline}
  \label{eq:elbo-z}
  \mathcal{F}\left [ \bm{\phi}(\mathbf{y} \mid \bm{\phi}), \bm{\theta} \right ] = \mathbb{E}_{\mathcal{N}(\mathbf{z} \mid 0, \mathbf{I}_N)} [ \log p(\mathcal{D}_s \mid \mathbf{y})]\\
  + \mathbb{E}_{\mathcal{N}(\mathbf{z} \mid 0, \mathbf{I}_N)} [\log p(\mathbf{y} \mid \bm{\theta})] + \log \det \mathbf{R}_q + \mathcal{H} [ \mathcal{N}(\mathbf{z} | 0, \mathbf{I}_N) ],
\end{multline}
where $\mathbf{y} = \bm{\mu}_q + \mathbf{R}_q \mathbf{z}$, and $\mathcal{H}[ \mathcal{N}(\mathbf{z} \mid 0, \mathbf{I}_N) ] = N (1 + \log 2 \pi) / 2$ is the differential entropy of the standard multivariate normal.
We then define the following unbiased estimate of the ELBO,
\begin{equation}
  \label{eq:elbo-z-est}
  f(\mathbf{z} ; \bm{\phi}, \bm{\theta}) = \log p(\mathcal{D}_s \mid \mathbf{y}) + \log \det \mathbf{R}_q  - \frac{1}{2} \left [%
    \mathbf{y}^{\top} \mathbf{C}^{-1}_p \mathbf{y} + \log \det \mathbf{C}_p - N%
  \right ],
\end{equation}
with gradients
\begin{align}
  \label{eq:elbo-z-est-grad-muq}
  \nabla_{\bm{\mu}_q} f(\mathbf{z} ; \bm{\phi}, \bm{\theta}) &= \nabla_{\mathbf{y}} \log p(\mathcal{D}_s \mid \mathbf{y}) - \mathbf{C}^{-1}_p \mathbf{y}, \\
  \label{eq:elbo-z-est-grad-Rq}
  \nabla_{\mathbf{R}_q} f(\mathbf{z} ; \bm{\phi}, \bm{\theta}) &= \left [ \nabla_{\mathbf{y}} \log p(\mathcal{D}_s \mid \mathbf{y}) \right ] \mathbf{z}^{\top} - \mathbf{C}^{-1}_p \mathbf{y} \mathbf{z}^{\top} + \left ( \mathbf{R}^{-1}_q \right )^{\top}, \\
  \label{eq:elbo-z-est-grad-theta}
  \frac{\partial}{\partial \theta_i} f(\mathbf{z} ; \bm{\phi}, \bm{\theta}) &= \frac{1}{2} \mathbf{y}^{\top} \mathbf{C}^{-1}_p \frac{\partial \mathbf{C}_p}{\partial \theta_i} \mathbf{C}^{-1}_p \mathbf{y} - \frac{1}{2} \operatorname{tr} \left ( \mathbf{C}^{-1}_p \frac{\partial \mathbf{C}_p}{\partial \theta_i} \right ),
\end{align}
where again $\mathbf{y} = \bm{\mu}_q + \mathbf{R}_q \mathbf{z}$.
The details of the derivations of~\eqref{eq:elbo-z-est-grad-muq}--\eqref{eq:elbo-z-est-grad-theta} are presented in~\ref{sec:gaussian-backprop}.
It can be verified that $\mathbb{E}_{\mathcal{N}(\mathbf{z} \mid 0, \mathbf{I}_N)}[f(\mathbf{z}; \bm{\phi}, \bm{\theta})] = \mathcal{F} \left [ q(\mathbf{y} \mid \bm{\phi}), \bm{\theta} \right ]$, so that the estimates are unbiased.

The variance of the estimate~\eqref{eq:elbo-z-est} and its gradients can be reduced by using the simple Monte Carlo (MC) or batch estimate and the corresponding gradients
\begin{gather}
  \label{eq:elbo-z-est-batch}
  f_n(\bm{\phi}, \bm{\theta}) = \frac{1}{n} \sum^n_{k = 1} f \left ( \mathbf{z}^{(k)}; \bm{\phi}, \bm{\theta} \right ), \quad \mathbf{z}^{(k)} \sim \mathcal{N}(0, \mathbf{I}_N),\\
  \label{eq:elbo-z-est-batch-grad}
  \nabla_{(\cdot)} f_n(\bm{\phi}, \bm{\theta}) = \frac{1}{n} \sum^n_{k = 1} \nabla_{(\cdot)} f(\mathbf{z}^{(k)}; \bm{\phi}, \bm{\theta}),
\end{gather}
where $n$ is the size of the batch.
The variance of the batch estimate~\eqref{eq:elbo-z-est-batch} is lower by a factor of $n$, but requires computing the gradients of the log-likelihood $n$ times.

Our DSVI algorithm is summarized in Algorithm~\ref{alg:dsvi}.
The stochastic gradient ascent algorithm with adaptive step-size sequence, proposed by \cite{kucukelbir-2017-automatic}, is reproduced in~\ref{sec:as3} for completeness.

\begin{algorithm}
  \caption{Doubly stochastic variational inference}
  \label{alg:dsvi}
  \begin{algorithmic}[0]
    \Require $\bm{\phi}^{(0)}$, $\bm{\theta}^{(0)}$, $\mathbf{C}_p(\bm{\theta})$, $\log p(\mathcal{D}_s \mid \mathbf{y})$%
    \State $j \leftarrow 0$%
    \Repeat%
    \State Sample $n$ realizations of $\mathbf{z} \sim \mathcal{N}(0, \mathbf{I}_N)$%
    \State Compute $\nabla_{\bm{\phi}} f_n(\bm{\phi}^{(j)}, \bm{\theta}^{(j)})$ and $\nabla_{\bm{\theta}} f_n(\bm{\phi}^{(j)}, \bm{\theta}^{(j)})$ using \eqref{eq:elbo-z-est-grad-muq}--\eqref{eq:elbo-z-est-grad-theta} and \eqref{eq:elbo-z-est-batch-grad}%
    \State Calculate step-size vectors $\rho^{(j)}_{\bm{\phi}}$ and $\rho^{(j)}_{\bm{\theta}}$ using \eqref{eq:as3-rho} and \eqref{eq:as3-s}
    \State $\bm{\phi}^{(j + 1)} \leftarrow \bm{\phi}^{(j)} + \bm{\rho}^{(j)}_{\bm{\phi}} \circ \nabla_{\bm{\phi}} f_n(\bm{\phi}^{(j)}, \bm{\theta}^{(j)})$ \eqref{eq:as3-phi-update}
    \State $\bm{\theta}^{(j + 1)} \leftarrow \bm{\theta}^{(j)} + \bm{\rho}^{(j)}_{\bm{\theta}} \circ \nabla_{\bm{\theta}} f_n(\bm{\phi}^{(j)}, \bm{\theta}^{(j)})$ \eqref{eq:as3-theta-update}
    \Until{Convergence}
  \end{algorithmic}
\end{algorithm}

\subsection{Parameterization of the factor matrix $\mathbf{R}_q$}
\label{sec:cholesky-factor-patterns}

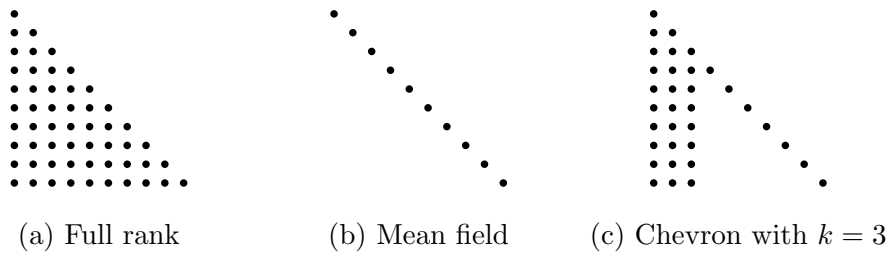
\begin{figure}[tbhp]
  \captionsetup[sub]{skip=12pt}
  \centering
  \begin{subfigure}[b]{0.3\textwidth}
    \centering
    \begin{tikzpicture}[x=0.25cm, y=0.25cm]
      \foreach \i in {0,...,9}{%
        \foreach \j in {0,...,9}{%
          \pgfplotstablegetelem{\i}{\j}\of\cfpfull%
          \ifnum\pgfplotsretval=0\relax\else%
          \node[circle, minimum size=0.1cm, inner sep=0pt, fill=black] at (\j, -\i) {};%
          \fi%
        };%
      };%
    \end{tikzpicture}
    \caption{Full rank}
    \label{fig:cholesky-factor-pattern-full}
  \end{subfigure}
  \begin{subfigure}[b]{0.3\textwidth}
    \centering
    \begin{tikzpicture}[x=0.25cm, y=0.25cm]
      \foreach \i in {0,...,9}{%
        \foreach \j in {0,...,9}{%
          \pgfplotstablegetelem{\i}{\j}\of\cfpdiag%
          \ifnum\pgfplotsretval=0\relax\else%
          \node[circle, minimum size=0.1cm, inner sep=0pt, fill=black] at (\j, -\i) {};%
          \fi%
        };%
      };%
    \end{tikzpicture}
    \caption{Mean field}
    \label{fig:cholesky-factor-pattern-diag}
  \end{subfigure}
  \begin{subfigure}[b]{0.3\textwidth}
    \centering
    \begin{tikzpicture}[x=0.25cm, y=0.25cm]
      \foreach \i in {0,...,9}{%
        \foreach \j in {0,...,9}{%
          \pgfplotstablegetelem{\i}{\j}\of\cfpchev%
          \ifnum\pgfplotsretval=0\relax\else%
          \node[circle, minimum size=0.1cm, inner sep=0pt, fill=black] at (\j, -\i) {};%
          \fi%
        };%
      };%
    \end{tikzpicture}
    \caption{Chevron with $k = 3$}
    \label{fig:cholesky-factor-pattern-chev}
  \end{subfigure}
  \caption{Parameterization of the factor matrix $\mathbf{R}_q$}
  \label{fig:cholesky-factor-patterns}
\end{figure}

It remains to discuss the parameterization of the factor $\mathbf{R}_q$.
In this manuscript, we consider three alternatives: a full parameterization, the so-called \emph{mean field} parameterization, and a constrained Chevron parameterization.
The sparsity patterns of these parameterizations are shown in~Figure~\ref{fig:cholesky-factor-patterns}.

In the \emph{full rank} parameterization~\cite{kucukelbir-2017-automatic}, we take $\mathbf{R}_q$ to be the non-unique Cholesky factor, that is, a $N \times N$ lower triangular matrix with unconstrained entries (Figure~\ref{fig:cholesky-factor-pattern-full}).
In this case, we have $\bm{\phi} \in \mathbb{R}^{N + N(N + 1) / 2}$.
The number of variational parameters is therefore $O(N^2)$, which may render their optimization difficult.
In order to address this challenge, we can employ the mean field and Chevron parameterizations, which result in a total number of variational parameters that is linear on $N$.

In the mean field parameterization, we take $\mathbf{R}_q$ to be a strictly positive diagonal matrix, i.e., $\mathbf{R}_q = \operatorname{diag}[\exp(\bm{\omega}_q)]$, with $\bm{\omega}_q \in \mathbb{R}^N$ (Figure~\ref{fig:cholesky-factor-pattern-diag}) and $\exp(\cdot)$ understood as element-wise.
This parameterization assumes that the variational density covariance is diagonal, and the exponential ensures that the non-zero entries of $\mathbf{R}_q$ are strictly positive.
In this case, we have $\bm{\phi} \equiv \{ \bm{\mu}_q, \bm{\omega}_q \} \in \mathbb{R}^{2 N}$.
The gradient of $f(\mathbf{z}; \bm{\phi}, \bm{\theta})$ with respect to $\bm{\omega}_q$ is given by
\begin{equation}
  \label{eq:elbo-z-est-grad-omegaq}
  \nabla_{\bm{\omega}_q} f(\mathbf{z}; \bm{\phi}, \bm{\theta}) = \left [ - \nabla_{\mathbf{y}} \log p(\mathcal{D}_s \mid \mathbf{y}) + \mathbf{C}^{-1}_p y \right ] \circ \mathbf{z} \circ \exp(\bm{\omega}_q) - \mathbf{I}_N,
\end{equation}
where $\circ$ denotes element-wise product, and exponentiation is taken as element-wise.
The details of the derivation are presented in~\ref{sec:gaussian-backprop}.
This parameterization assumes that the posterior components of $\mathbf{y}$ are essentially uncorrelated and therefore cannot resolve the correlations of the true posterior, which are expected to be non-trivial for highly correlated prior covariance structures and for a small number of observations.
As a consequence, the mean field parameterization tends to underestimate the posterior variance~\cite{blei-2017-variational}.

Finally, the constrained Chevron parameterization is similar to the full parameterization, but we set entries below the diagonal and for column number larger than the Chevron parameter $k < N$ to zero (Figure~\ref{fig:cholesky-factor-pattern-chev}) \cite{challis-2013-gaussian}.
In this case, we have $\bm{\phi} \in \mathbb{R}^{N + (2 N - k)(k + 1) / 2}$.
The number of variational parameters for this parameterization is $O(N (k + 1))$, a reduction with respect to the full parameterization, while maintaining some degree of expressivity for capturing correlations of the true posterior.

%\subsection{Relation to previous work}
%\label{sec:dsvi-previous}

%
The DSVI-EB method follows the \emph{automatic differentiation variational inference} (ADVI) algorithm~\cite{kucukelbir-2017-automatic}, in which gradients of the joint probability $p(\mathcal{D}_s \mid \mathbf{y})$ with respect to variational parameters are computed using Gaussian backpropagation and reverse-mode automatic differentiation.
ADVI is formulated for the full Bayes case and implements the full and mean-field parameterizations of $\mathbf{R}_q$.
In comparison, our work is formulated for the empirical Bayes case, employs the adjoint method to compute gradients of physics solvers, and implements the constrained Chevron parameterization in addition to the full and mean-field parameterizations.

Two schemes are common in the literature for computing the gradients of the noisy ELBO estimate: the \textsc{reinforce} algorithm~\cite{willliams-1992-simple}, also known as the likelihood ratio method or the log-derivative trick, and Gaussian backpropagation~\cite{rezende-2014-stochastic}, also known as the reparameterization trick~\cite{kingma-2013-auto,titsias-2014-doubly}.
The \textsc{reinforce} algorithm employs gradients of the variational density with respect to its parameters, which is convenient as it only requires zero-order information of the physics model.
Unfortunately the \textsc{reinforce} estimates of the ELBO gradients are well-known to be of high variance and must be paired with a variance reduction technique~\cite{ranganath-2014-black}.
Gaussian backpropagation, on the other hand, results in lower-variance gradient estimates at the cost of requiring first-order information of the physics model.

An alternative formulation of VI is presented in~\cite{tsilifis-2016-computationally}, where the authors employ mixtures of diagonal multivariate Gaussian densities as the variational posterior, and approximate the ELBO using a second order Taylor expansion around the mean of each mixture component.
The mean, diagonal covariance and mixture weights are estimated by maximizing the ELBO via coordinate ascent.
This entirely deterministic approach is formulated for the inference problem and does not consider optimization over prior hyperparameters.
A similar approach is also presented in~\cite{friston-2007-variational} in the context of empirical Bayes.

\section{Computational cost}
\label{sec:cost}

In this section, we discuss the computational effort of the Laplace-EM and DSVI-EB algorithms.
We compute the gradient and the Hessian of the likelihood via the discrete adjoint method (see~\ref{sec:da} for details).
Note that the Laplace-EM method requires both gradients and Hessians, while the DSVI-EB method only requires gradients.
For the physics constraint $\mathbf{L}(\mathbf{u}, \mathbf{y}) = 0$, the computation of the gradient requires the solution of one (linear) backward sensitivity problem of size $M \times M$.
Similarly, the computation of the Hessian requires one backward sensitivity problem and $N$ forward sensitivity problems, each of size $M \times M$.
For the following discussion we assume that the cost of each forward and backward sensitivity problem is of order $O(M^{\gamma})$, $\gamma > 1$.

For the Laplace-EM algorithm we discuss the cost per each EM cycle.
Each E-step requires one Cholesky factorization of $\mathbf{C}_p(\bm{\theta}^{(j)})$, of cost $N^3$, the solution of~\eqref{eq:laplace-mean} via gradient-based optimization, and the computation of the Hessian.
Therefore, the cost of each E-step is $O(\max( M^{\gamma}, N^3))$.
Each iteration of the M-step requires one Cholesky factorization of $\mathbf{C}_p(\bm{\theta})$.
Therefore, the total cost of each EM cycle is again $O(\max( M^{\gamma}, N^3))$.

For the DSVI-EB algorithm, each iteration requires evaluating $n$ gradients and one Cholesky factorization of $\mathbf{C}_p$.
If $n$ is chosen independent of $M$, we have that the total cost per iteration is also $O(\max( M^{\gamma}, N^3))$.

Finally, in general we expect the number of iterations for each $E$- and $M$-step, and the number of EM cycles and DSVI iterations, to increase with increasing $N$.
The analysis of how said numbers scale with $N$ is beyond the scope of this manuscript.

\section{Numerical experiments}
\label{sec:numexp}

In this section, we present the application of the proposed approximate inference algorithms to the identification of the diffusion coefficient in diffusion equations.
In particular, we are interested in identifying the diffusion coefficient $k(\mathbf{x}, u)$ of the homogeneous diffusion equation $\nabla \cdot (k(\mathbf{x}, u) \nabla u) = 0$ in $\Omega \subset \mathbb{R}^d$, from both measurements of the diffusion coefficient and of the state $u$.
For the linear case ($k \equiv k(\mathbf{x})$), the diffusion equation models phenomena such as stationary heat transfer and Darcy flow.
For the nonlinear case ($k \equiv k(u)$), one recovers the so-called Richards equation for horizontal flows in unsaturated porous media.

\subsection{Linear diffusion problem}
\label{sec:darcy}

We consider the one-dimensional diffusion equation with Dirichlet boundary conditions
\begin{gather}
  \label{eq:darcy}
  \frac{\partial}{\partial x} \left [ k(x) \frac{\partial}{\partial x} u(x) \right ] = 0, \quad x \in [0, 1],\\
  \label{eq:darcy-bc}
  u(0) = u_{\mathrm{L}}, \quad u(1) = u_{\mathrm{R}},
\end{gather}
where $u \colon [0, 1] \to \mathbb{R}$ is the state and $k \colon [0, 1] \to \mathbb{R}^+$ is the diffusion coefficient.
The state is discretized into $M$ degrees of freedom $u_i$ organized into the vector $\mathbf{u} \in \mathbb{R}^M$.
The diffusion coefficient is discretized into $N$ degrees of freedom $k_i = \exp y_i$ corresponding to $N$ spatial coordinates $\{ x_i \}^N_{i = 1}$, with the $y_i$ organized into the vector $\mathbf{y} \in \mathbb{R}^N$.
The discretized problem~\eqref{eq:darcy} and \eqref{eq:darcy-bc} is of algebraic form $\mathbf{L}(\mathbf{u}, \mathbf{y}) \equiv \mathbf{S}(\mathbf{y}) \mathbf{u} - \mathbf{b}(\mathbf{y}) = 0$, where $\mathbf{S} \colon \mathbb{R}^N \to \mathbb{R}^{M \times M}$ and $\mathbf{b} \colon \mathbb{R}^N \to \mathbb{R}^M$.
In  \eqref{eq:darcy}, $\mathbf{y}$ can only be identified from measurements of $u$ up to an additive constant~\cite{stuart_2010}, and measurements of $\mathbf{y}$ are required to estimate it uniquely. 

\begin{figure}[tbhp]
  \centering
  \begin{subfigure}[b]{0.45\textwidth}
    \includegraphics[width=\textwidth]{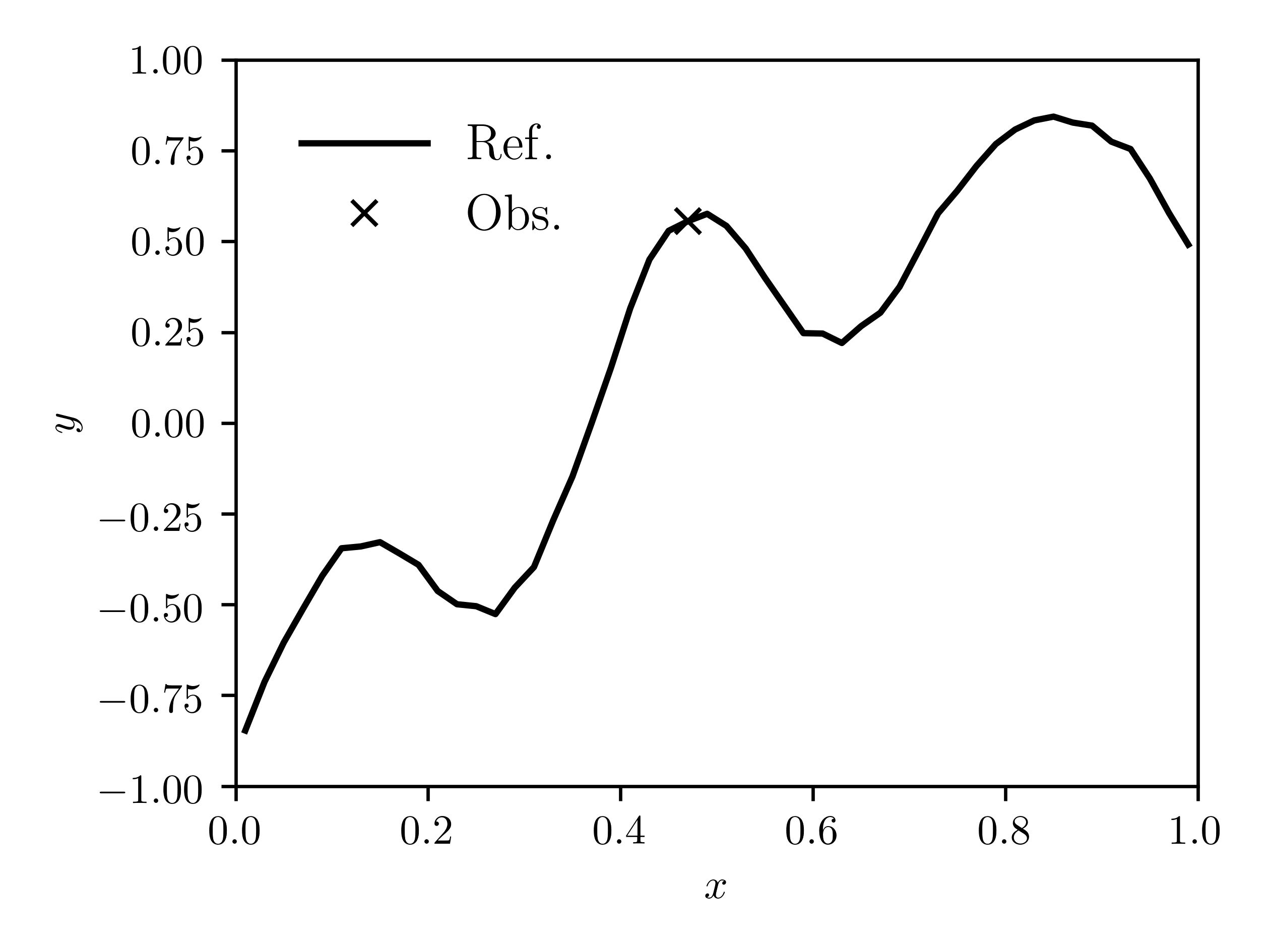}
    \caption{}
    \label{fig:darcy-yref}
  \end{subfigure}
    \begin{subfigure}[b]{0.45\textwidth}
    \includegraphics[width=\textwidth]{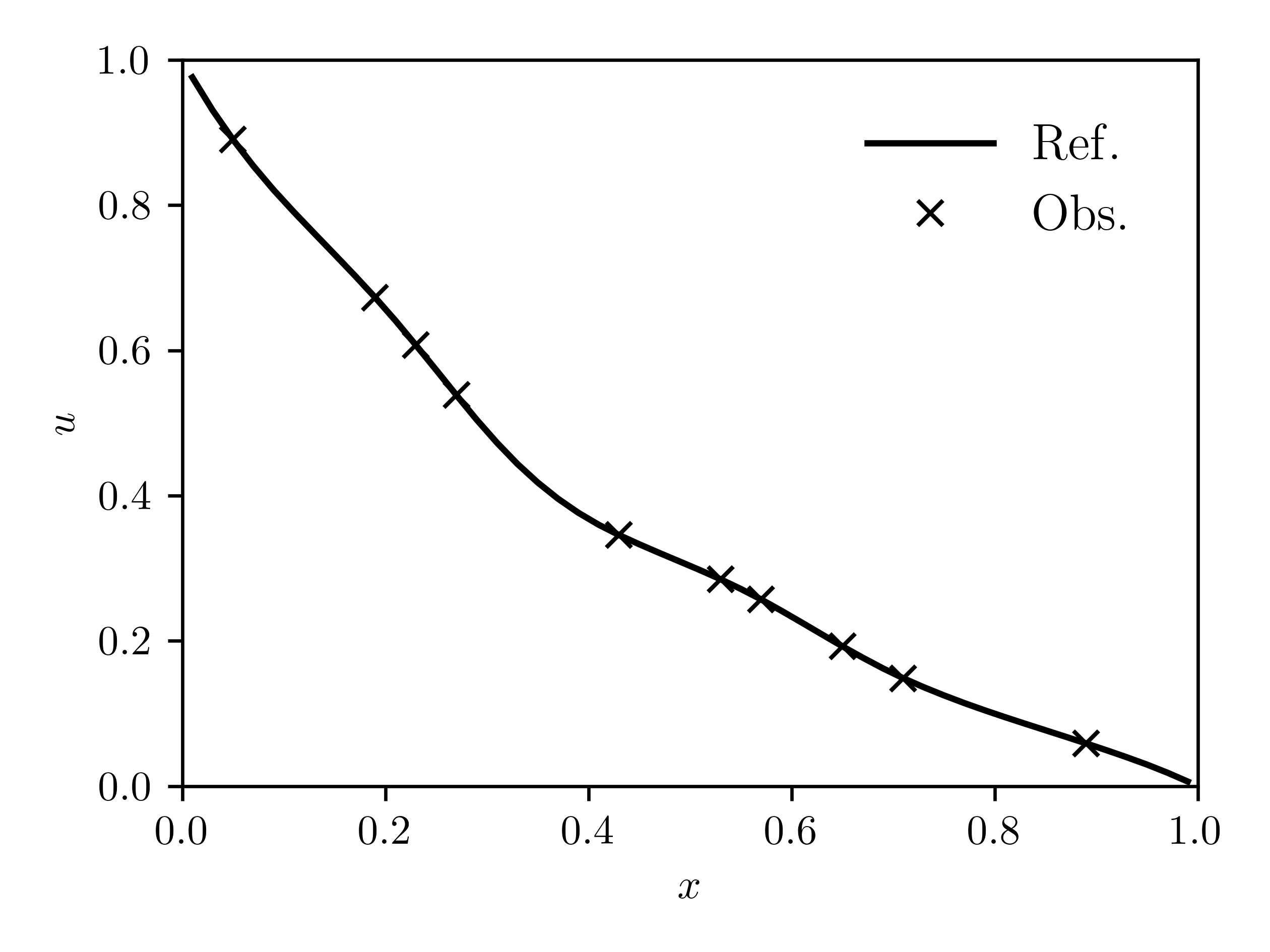}
    \caption{}
    \label{fig:darcy-uref}
  \end{subfigure}
  \caption{Reference diffusion coefficient and state fields (continuous lines), and observations (crosses), for the one-dimensional linear diffusion problem}
  \label{fig:darcy-ref}
\end{figure}

We apply the presented model inversion algorithms to estimating a synthetic diffusion coefficient from $10$ measurements of the state $u$ and one measurement of the log-diffusion coefficient $y$.
The reference values of $\mathbf{y}$ and $\mathbf{u}$ and the corresponding observations are shown in~Figure~\ref{fig:darcy-ref}.
The reference $\mathbf{y}$ is taken as a realization of the zero-mean GP with squared exponential covariance
\begin{equation}
  \label{eq:darcy-prior-covar}
  C(x, x' \mid \bm{\theta}) = \sigma^2 \exp \left [ -\left (x - x' \right)^2 / 2 \lambda^2 \right ] + \sigma^2_n 1_{x = x'},
\end{equation}
where $\bm{\theta} \equiv (\sigma, \lambda)$, and $\sigma_n$ is set to $\num{1e-2}$.
We refer to $\sigma$ and $\lambda$ as the standard deviation and correlation length, respectively, of the prior covariance.
The reference values of $\bm{\theta}$ are presented in~Table~\ref{tab:darcy-theta}.
The $\mathbf{y}$ and $\mathbf{u}$ observations are taken at randomly selected degrees of freedom, with observation error standard deviations $\sigma_{us} = \sigma_{ys} = \num{1e-3}$.
Finally, the boundary conditions are set to $u_{\mathrm{L}} = 1.0$ and $u_{\mathrm{R}} = 0.0$, and the numbers $M$ and $N$ of degrees of freedom is set to $\num{50}$.

\subsubsection{Empirical Bayesian inference}
\label{sec:darcy-eb}

\begin{figure}[tbhp]
  \centering
  \begin{subfigure}[b]{0.45\textwidth}
    \includegraphics[width=\textwidth]{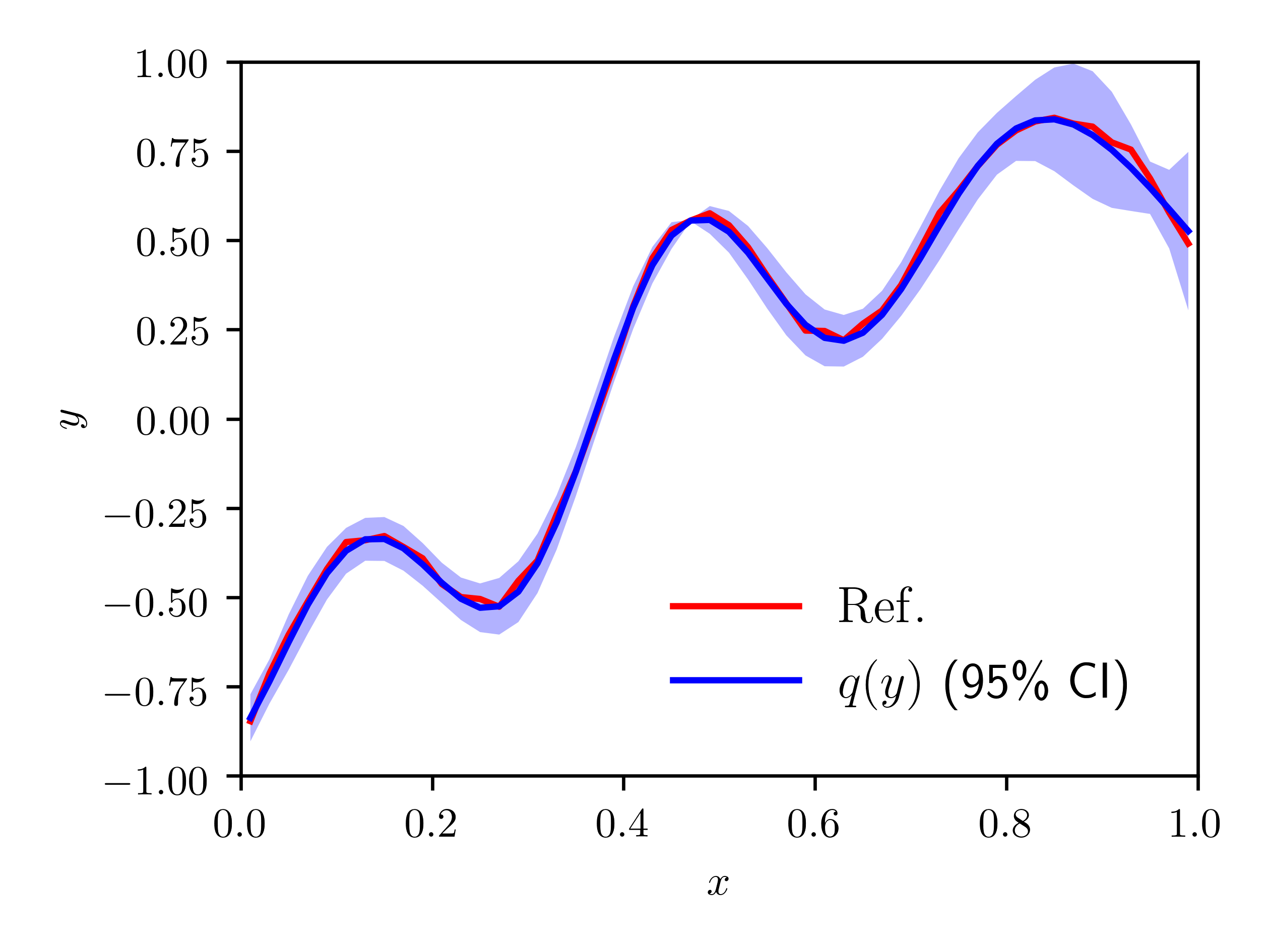}
    \caption{Laplace-EM}
    \label{fig:darcy-laplace-em}
  \end{subfigure}
    \begin{subfigure}[b]{0.45\textwidth}
    \includegraphics[width=\textwidth]{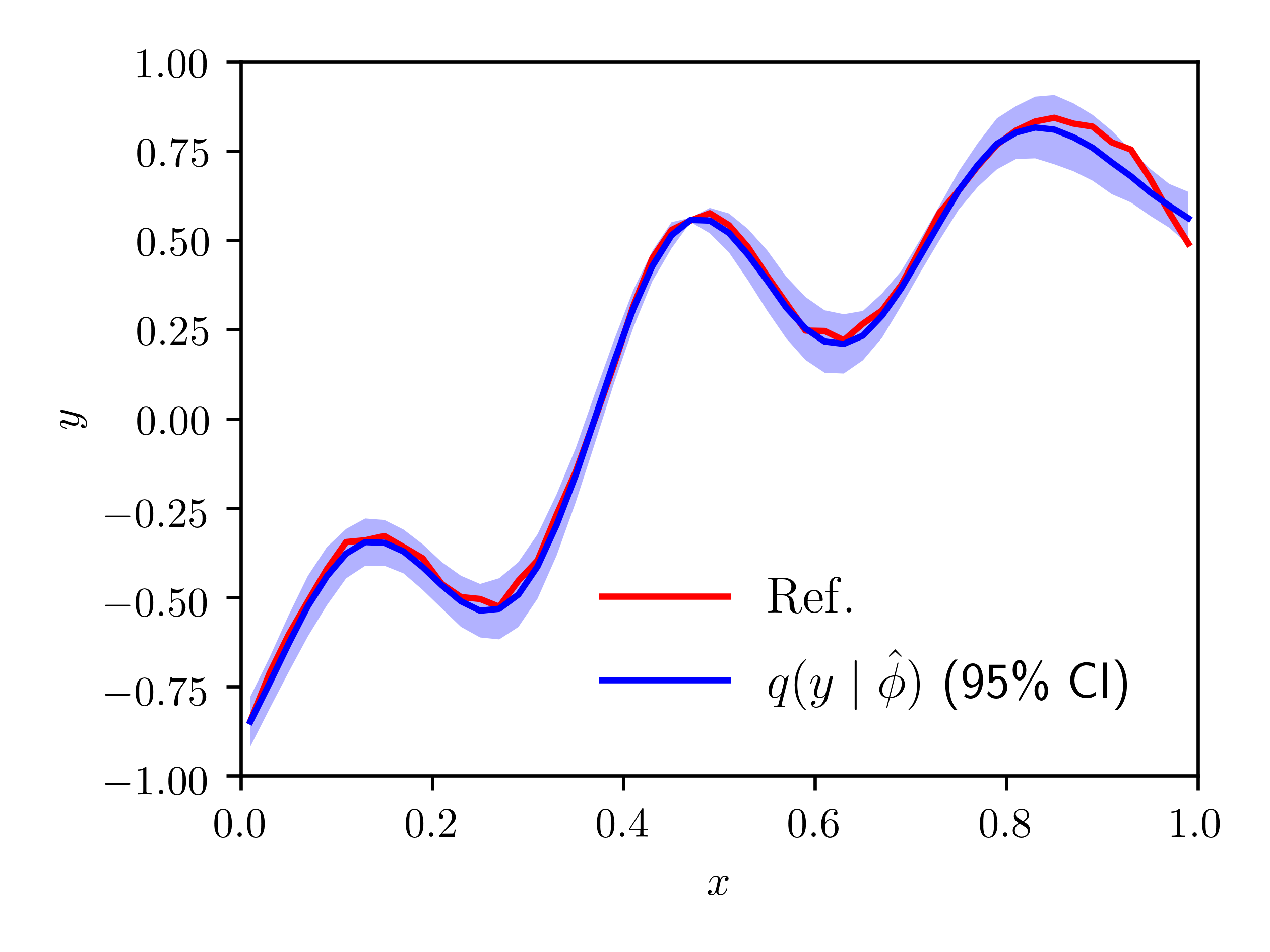}
    \caption{DSVI with Chevron factor $k = 20$}
    \label{fig:darcy-dsvi-eb}
  \end{subfigure}
  \caption{Reference and estimated diffusion coefficient for the one-dimensional linear diffusion problem}
  \label{fig:darcy-yest}
\end{figure}

Figure~\ref{fig:darcy-yest} shows the estimated diffusion coefficient, together with the $95\%$ confidence intervals centered around the posterior mean, computed using the Laplace-EM algorithm and DSVI-EB with Chevron parameterization and $k = 20$.
It can be seen that both methods provide accurate estimates of the reference field, with the reference field falling inside the confidence interval of the estimates (with localized exceptions for DSVI-EB with Chevron parameterization in the vicinity of the $x = 1.0$ boundary, as shown in Figure~\ref{fig:darcy-dsvi-eb}).

The estimated prior hyperparameters are presented in~Table~\ref{tab:darcy-theta}, together with a simple MC estimate of the ELBO, $\hat{\mathcal{F}}$, computed using $\num{1e4}$ realizations of the estimated posterior.
It can be seen that hyperparameter estimates are very similar for the Laplace-EM method and the DSVI method.
In particular, estimates of the correlation length are close to reference values, while the standard deviation is underestimated across all methods.
Estimates are also similar for the different factor matrix parameterizations on the DSVI method, with the exception of the full rank parameterization that resulted in more pronounced underestimation of both the standard deviation and correlation length.
In terms of the ELBO, the Laplace-EM method results in the highest value, followed by the DSVI method with full rank factor parameterization.
This indicates that full rank representations of the covariance matrix of the estimated posterior density result in better estimates of the true posterior, whereas reduced representations such as Chevron and mean field are less accurate.
In practice, it can be seen in Figure~\ref{fig:darcy-dsvi-eb} that the reduced representation is less capable of resolving the uncertainty of the $y$ estimate in the vicinity of the $x = 1.0$ boundary.
Nevertheless, reduced representations are not worse than fuller representations for estimating prior hyperparameters.

\begin{table}[tbhp]
  \centering
  \begin{tabular}{lllcc}
    \toprule
    & & & \multicolumn{2}{c}{Hyperparameters} \\
    & & $\hat{\mathcal{F}}$ & $\sigma$ & $\lambda$\\
    \midrule
    \multicolumn{2}{l}{Reference} & & \num{1.000} & \num{0.150}\\
    \multicolumn{2}{l}{Laplace-EM} & \num{-37.37 \pm 0.05} & \num{0.608} & \num{0.132}\\
    DSVI & Full rank & \num{-43.38 \pm 0.08} & \num{0.551} & \num{0.120}\\
    & Chevron $k = 20$ & \num{-49.54 \pm 0.09}  & \num{0.653} & \num{0.144}\\
    & Chevron $k = 10$ & \num{-50.47 \pm 0.09}  & \num{0.672} & \num{0.147}\\
    & Chevron $k = 5$  & \num{-47.85 \pm 0.06}  & \num{0.681} & \num{0.148}\\
    & Mean field       & \num{-49.34 \pm 0.06}  & \num{0.687} & \num{0.151}\\
    \bottomrule
  \end{tabular}
  \caption{Reference and estimated hyperparameters, and simple MC estimate of the ELBO, for the one-dimensional linear-diffusion problem}
  \label{tab:darcy-theta}
\end{table}

\subsubsection{Comparison against MCMC}
\label{sec:darcy-comparison}

We proceed to evaluate the accuracy of the proposed inference algorithms at approximating the posterior density $p(\mathbf{y} \mid \mathcal{D}_s, \bm{\theta})$.
For this purpose, 
  we employ MCMC simulation as the benchmark as it is known to converge to the exact posterior density.
  In order to restrict the focus to the approximation of the posterior density,
we set the prior hyperparameters to fixed values equal to the reference values, $\bm{\theta}_{\mathrm{ref}}$, and employ the Laplace-EM %
\footnote{%
  Note that in this context the Laplace-EM algorithm is reduced to the Laplace approximation for given $\bm{\theta}$ (e.g. a single E-step in the EM algorithm), but we will refer to the associated results as Laplace-EM results for the sake of convenience. 
}%
and DSVI algorithms to estimate the posterior $p(\mathbf{y} \mid \mathcal{D}_s, \bm{\theta}_{\mathrm{ref}})$.%
We compare the estimated posterior mean and standard deviation against the sample mean and standard deviation computed from $\num{1e4}$ MCMC realizations of the posterior generated using the No-U-Turn Sampler (NUTS)~\cite{hoffman-2014-nuts}.

\begin{figure}[tbhp]
  \centering
  \begin{subfigure}[b]{1.00\textwidth}
    \includegraphics[width=\textwidth]{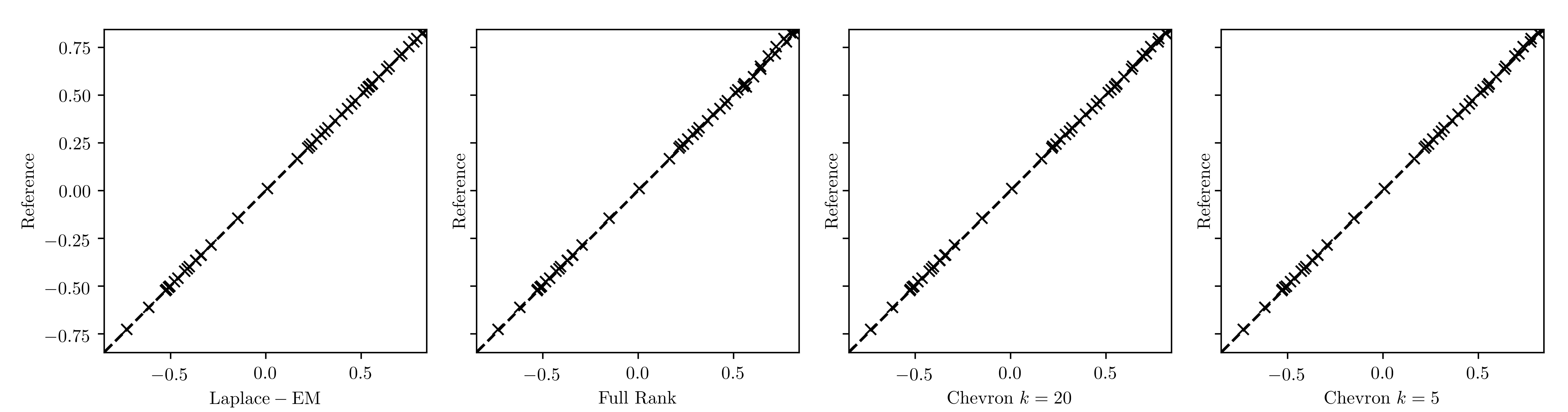}
    \caption{Mean}
    \label{fig:darcy-comparison-mean}
  \end{subfigure}
  \begin{subfigure}[b]{1.00\textwidth}
    \includegraphics[width=\textwidth]{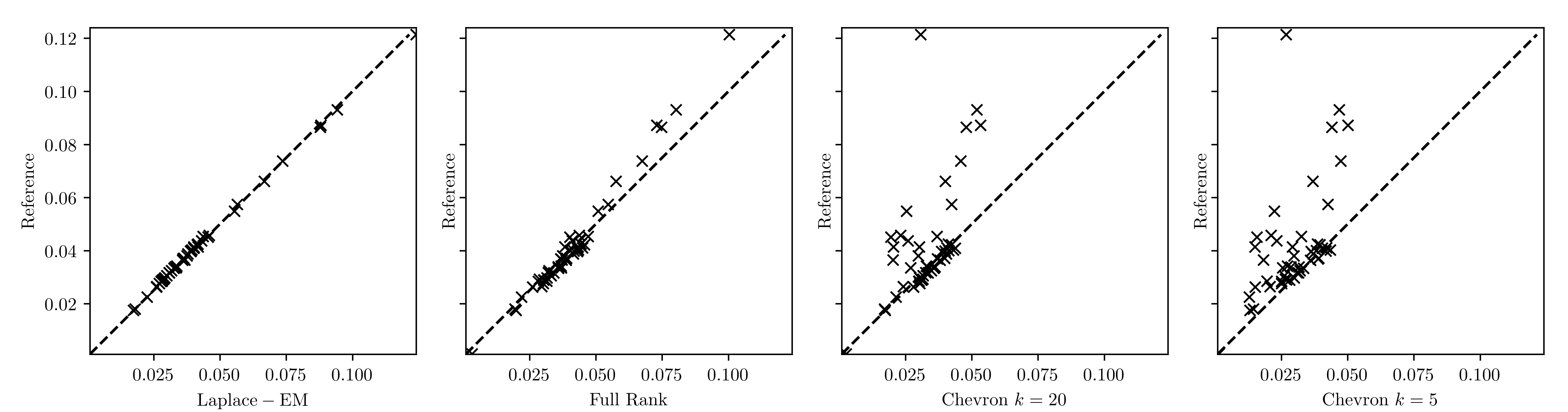}
    \caption{Standard deviation}
    \label{fig:darcy-comparison-stddev}
  \end{subfigure}
  \caption{Posterior mean and standard deviation estimated via approximate inference, compared against sample mean and standard deviation computed from MCMC realizations of the posterior (reference)}
  \label{fig:darcy-comparison}
\end{figure}

Figure~\ref{fig:darcy-comparison} presents a point-wise comparison against MCMC of the estimated mean and standard deviation computed using both Laplace-EM and DSVI with the full rank parameterization and the Chevron parameterization with $k = 20$ and $5$.
It can be seen that all estimates of the mean are very accurate, which indicates that the presented approximate inference algorithms provide accurate estimates of the mean of multimodal posterior densities.
This result is expected for the Laplace-EM algorithm where the estimated posterior mean is set to the MAP, but for the DSVI algorithm this is less of a given.

For the standard deviation, the Laplace-EM method provides the most accurate estimates, followed by DSVI with the full rank parameterization.
This reinforces the conclusion drawn previously that full rank representations of the covariance lead to better estimates of the true posterior.
Furthermore, it can be seen that the Chevron representation accurately resolves the bulk of point-wise standard deviation values (clustered at the bottom left of each plot in Figure~\ref{fig:darcy-comparison-stddev}) but leads to noticeable underestimation of the larger point-wise values (i.e. the top half of Figure~\ref{fig:darcy-comparison-stddev}).
The underestimation of the standard deviation is more pronounced for decreasing $k$, and is the most pronounced for the mean field parameterization (not shown), which as remarked previously tends to underestimate the variance of the posterior~\cite{blei-2017-variational}.

\begin{figure}[tbhp]
  \centering
  \begin{subfigure}[b]{0.45\textwidth}
    \includegraphics[width=\textwidth]{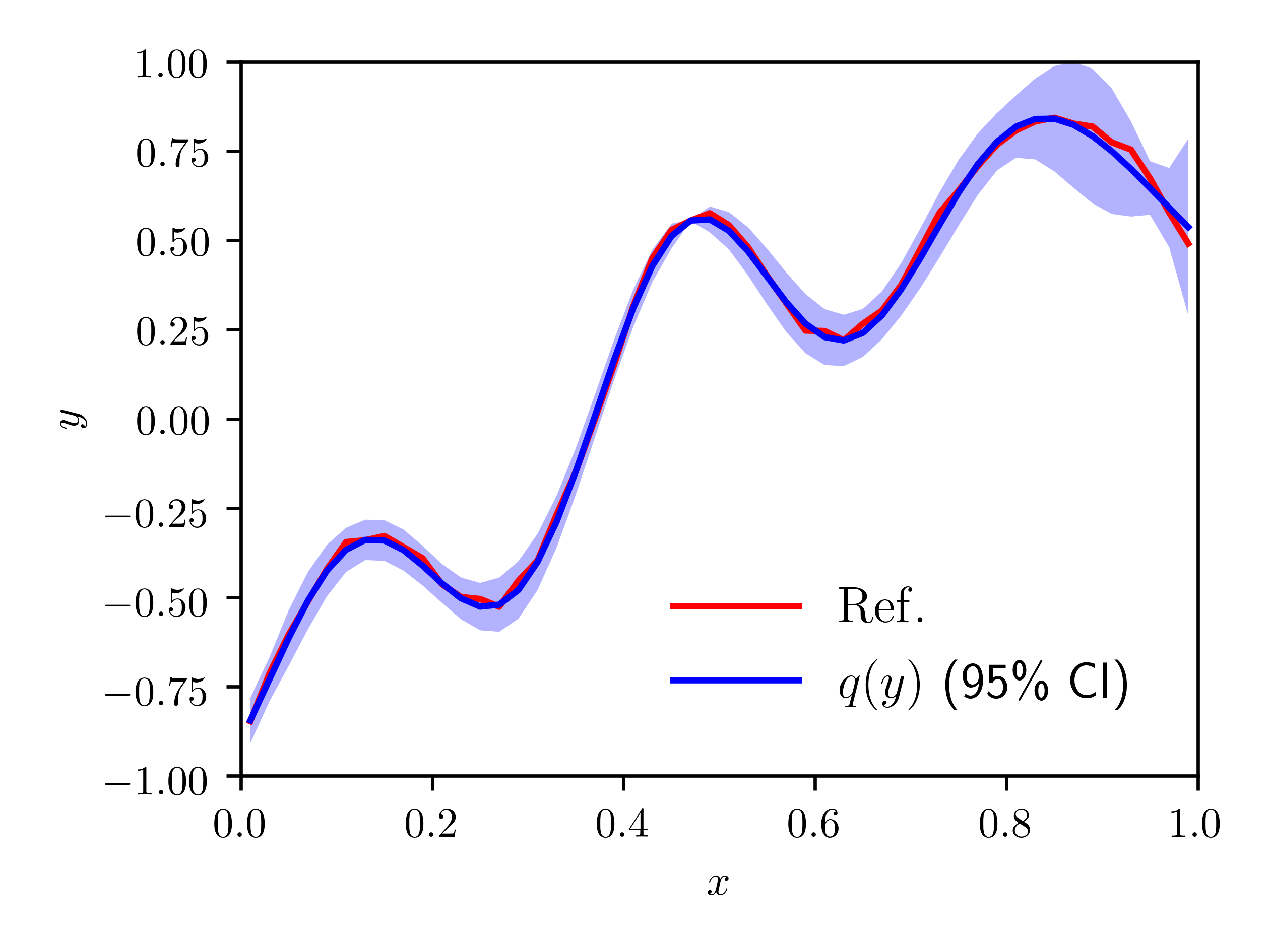}
    \caption{Laplace-EM}
    \label{fig:darcy-laplace-em-comparison}
  \end{subfigure}
    \begin{subfigure}[b]{0.45\textwidth}
    \includegraphics[width=\textwidth]{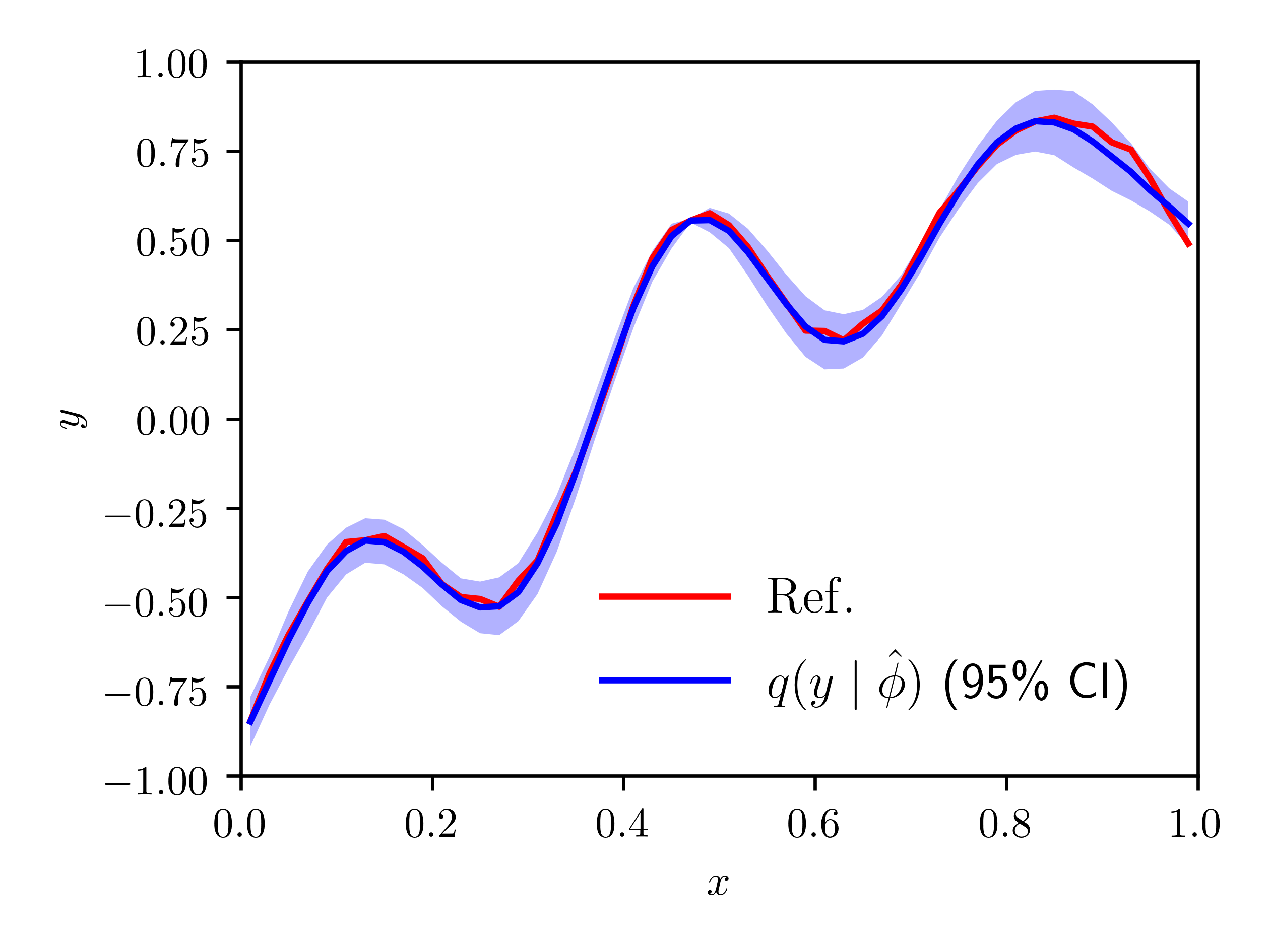}
    \caption{DSVI with Chevron factor $k = 20$}
    \label{fig:darcy-dsvi-eb-comparison}
  \end{subfigure}
  \caption{Estimated posterior mean and 95\% confidence interval computed from the estimated posterior variance, for the one-dimensional linear diffusion problem for fixed $\bm{\theta} = \bm{\theta}_{\mathrm{ref}}$}
  \label{fig:darcy-yest-comparison}
\end{figure}

Finally, in~Figure~\ref{fig:darcy-yest-comparison} we present the posterior mean and variance for Laplace-EM and DSVI with Chevron parameterization and $k = 20$, obtained for fixed $\bm{\theta} = \bm{\theta}_{\mathrm{ref}}$.
Comparing~Figure~\ref{fig:darcy-yest} against~Figure~\ref{fig:darcy-yest-comparison} reveals that even though the empirical Bayes estimation procedure results in a standard deviation estimate lower than the reference value (see Table~\ref{tab:darcy-theta}), the posterior density with empirical Bayes hyperparameter estimates, $p(\mathbf{y} \mid \mathcal{D}_s, \hat{\bm{\theta}})$, is a good approximate to the posterior density with reference hyperparameters, $p(\mathbf{y} \mid \mathcal{D}_s, \bm{\theta}_{\mathrm{ref}})$.

\subsection{Nonlinear diffusion problem}
\label{sec:richards}

We consider the one-dimensional nonlinear diffusion equation with Dirichlet boundary conditions
\begin{gather}
  \label{eq:richards}
  \frac{\partial}{\partial x} \left [ k(u(x)) \frac{\partial}{\partial x} u(x) \right ] = 0, \quad x \in [0, 1],\\
  \label{eq:richards-bc}
  u(0) = u_{\mathrm{L}}, \quad u(1) = u_{\mathrm{R}}, \quad u_L < u_R \leq 0,
\end{gather}
where $u \colon [0, 1] \to (-\infty, 0]$ is the state and $k \colon (\infty, 0] \to \mathbb{R}^+$ is the diffusion coefficient.
Similarly to Section~\ref{sec:darcy}, the state is discretized into $M$ degrees of freedom $u_i$ organized into the vector $\mathbf{u} \in \mathbb{R}^M$.
The diffusion coefficient function  is discretized into $N$ degrees of freedom $k_i = \exp y_i$ corresponding to $N$ values of $u$ over $[u_{\min}, 0]$ (where $u_{\mathrm{min}} < u_{\mathrm{L}}$), organized into the vector $\mathbf{y} \in \mathbb{R}^N$.
The discretized problem~\eqref{eq:richards} and \eqref{eq:richards-bc} is of the algebraic form $\mathbf{L}(\mathbf{u}, \mathbf{y}) \equiv \mathbf{S}(\mathbf{u}, \mathbf{y}) \mathbf{u} - \mathbf{b}(\mathbf{u}, \mathbf{y}) = 0$, where $\mathbf{S} \colon \mathbb{R}^M \times \mathbb{R}^N \to \mathbb{R}^{M \times M}$ and $\mathbf{b} \colon \mathbb{R}^M \times \mathbb{R}^N \to \mathbb{R}^M$.
Inspection of \eqref{eq:richards} reveals that $y(u) \equiv \log k(u)$ can only be identified over the range $[u_{\mathrm{L}}, u_{\mathrm{R}}]$ and up to an additive constant%
\footnote{%
  This can be verified by introducing the Kirchhoff transformation $f(u) = \int^u_{u_{\min}} k(u) \, \mathrm{d} u$, with which \eqref{eq:richards} can be written as a linear equation on $f$.%
}.%
To disambiguate the estimate, we provide measurements of $y(u)$ at $u = u_{\min}$ and $u = 0$.

\begin{figure}[tbhp]
  \centering%
  \includegraphics[width=0.45\textwidth]{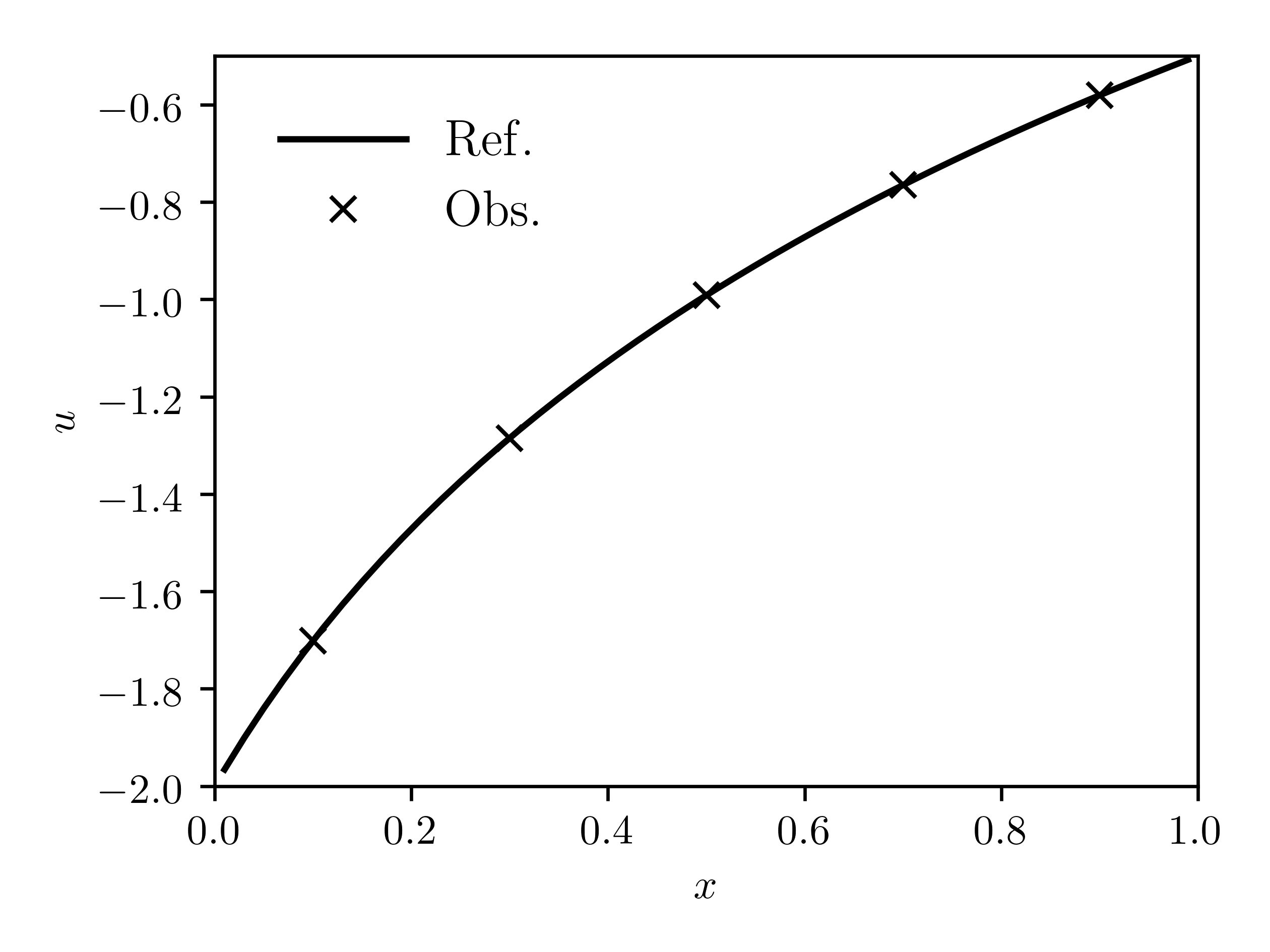}
  \caption{Reference state field (continuous lines), and observations (crosses), for the one-dimensional nonlinear diffusion problem}
  \label{fig:richards-uref}
\end{figure}

We apply the presented model inversion algorithms to estimating a function $k(u)$ from $5$ measurements of the state $u$ and 2 measurements of $y(u) \equiv \log k(u)$ at $u = u_{\min}$ and $u = 0$.
The reference diffusion coefficient is $k(u) = \exp u$ ($y(u) = u$).
The $\mathbf{u}$ observations are taken at randomly selected degrees of freedom, and are shown in Figure~\ref{fig:richards-uref}.
Observation error standard deviations $\sigma_{us}$ and $\sigma_{ys}$ are set to $\num{1e-2}$.
Boundary conditions are set to $u_{\mathrm{L}} = -2.0$ and $u_{\mathrm{R}} = -0.5$, and $u_{\min}$ is set to $-2.5$.
Finally, the numbers $M$ and $N$ are set to $50$ and $21$, respectively.

\begin{figure}[tbhp]
  \centering
  \begin{subfigure}[b]{0.45\textwidth}
    \includegraphics[width=\textwidth]{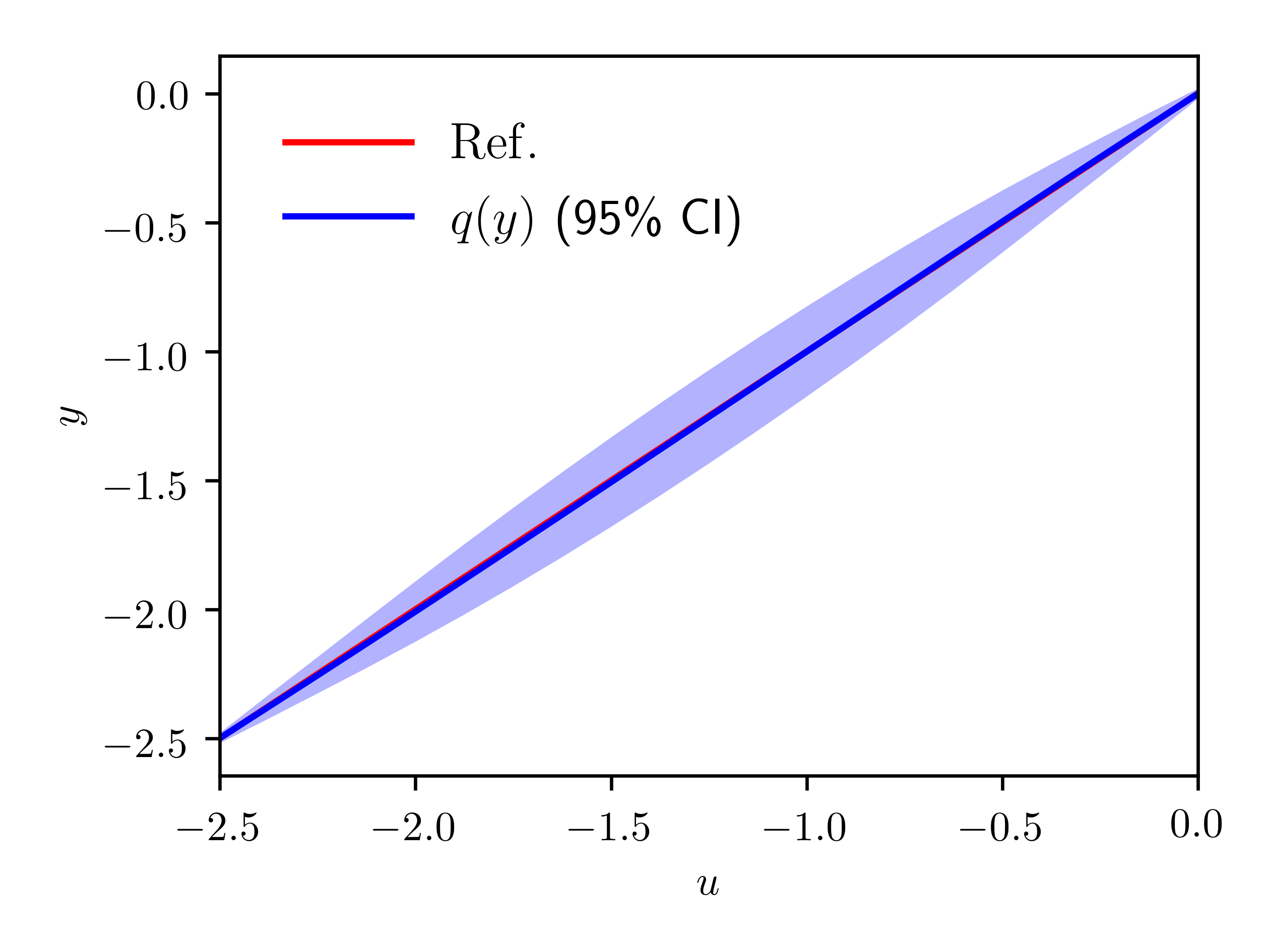}
    \caption{Laplace-EM}
    \label{fig:richards-laplace-em}
  \end{subfigure}
    \begin{subfigure}[b]{0.45\textwidth}
    \includegraphics[width=\textwidth]{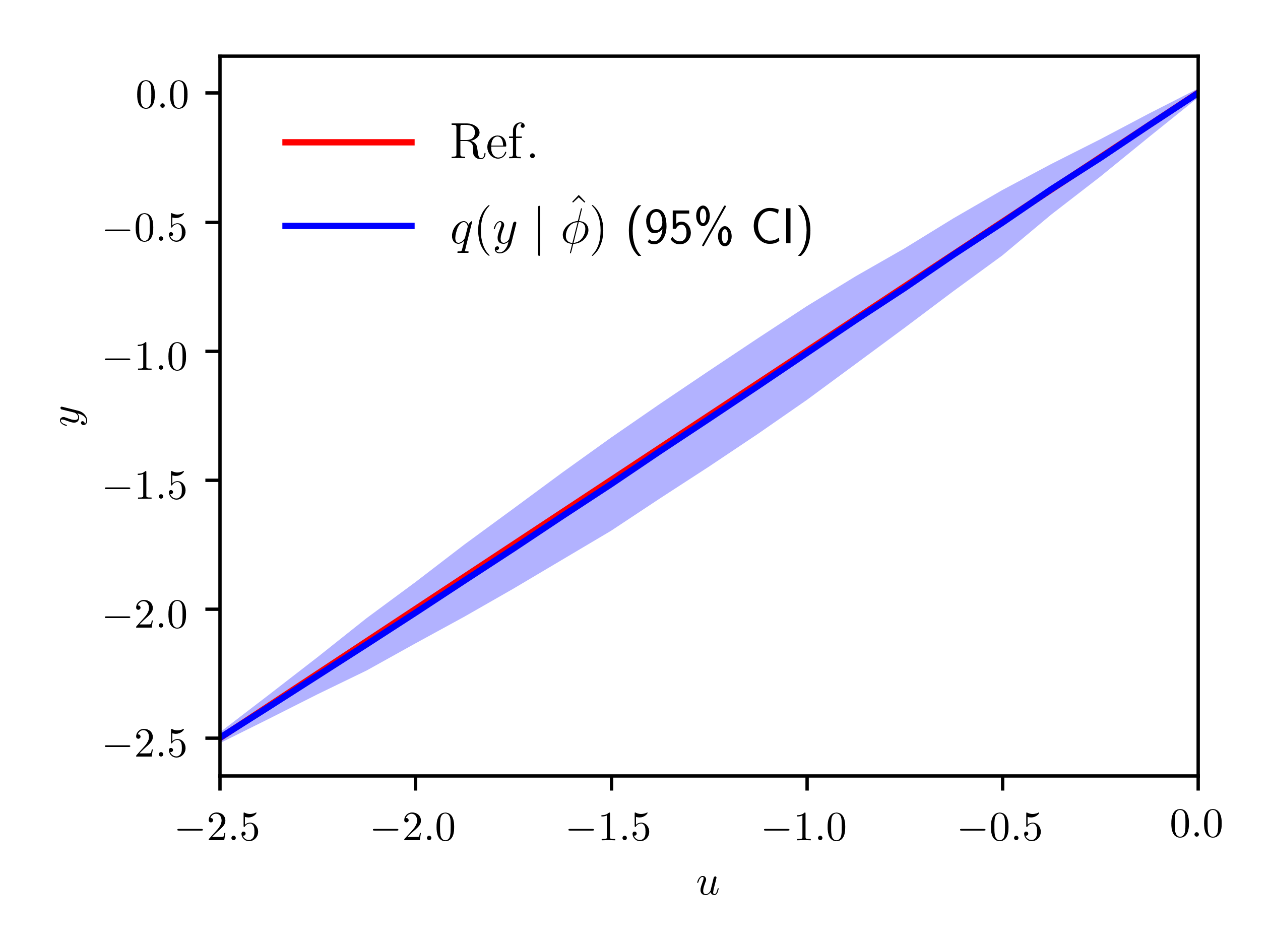}
    \caption{DSVI with Chevron factor, $k = 5$}
    \label{fig:richards-dsvi-eb}
  \end{subfigure}
  \caption{Estimated diffusion coefficient for the one-dimensional nonlinear diffusion problem}
  \label{fig:richards-yest}
\end{figure}

Figure~\ref{fig:richards-yest} shows the estimated diffusion coefficient using the proposed model inversion methods, together with the 95\% confidence intervals centered around the posterior mean.
Presented are the results for the Laplace-EM method and the DSVI-EB method with Chevron parameterization and $k = 5$.
  It can be seen that the estimated posterior mean and confidence intervals for both methods are nearly identical.
As prior covariance $C(u, u \mid \bm{\theta})$, we employ the squared exponential model~\eqref{eq:darcy-prior-covar} with $\sigma_n$ set to $\num{1e-2}$.
As in the linear case, both methods accurately estimate the reference function $y(u)$, and the reference function falls inside the 95\% confidence intervals provided by the estimated posterior covariance.

Table~\ref{tab:richards-theta} presents the estimated hyperparameters of the prior for the Laplace-EM and the DSVI-EB method, together with simple MC estimates of the ELBO computed using $\num{1e4}$ realizations of the corresponding estimated posterior densities.
It can be seen that estimated hyperparameters are different for the different methods (note that here we don't have reference values for the hyperparameters of the prior, as the reference $k(u)$ is not drawn from a GP model).
Nevertheless, it can be seen that both methods result in similar estimates of $y$.
In agreement with the linear case, the Laplace-EM and the DSVI-EB method with full rank parameterization result in the largest values of ELBO.
Additionally, it can be seen that the ELBO decreases with increasing sparsity of the posterior covariance factor parameterization (i.e. with decreasing Chevron factor $k$), being the lowest for the mean field parameterization.
This illustrates the compromise between the sparsity of the covariance factor and its expressive capacity for approximating the true posterior, %
that is, that less sparse covariance factors produce more accurate approximate posteriors.

\begin{table}[tbhp]
  \centering
  \begin{tabular}{lllcc}
    \toprule
    & & & \multicolumn{2}{c}{Hyperparameters} \\
    & & $\hat{\mathcal{F}}$ & $\sigma$ & $\lambda$\\
    \midrule
    \multicolumn{2}{l}{Laplace-EM} & \num{-12.68 \pm 0.03} & \num{4.650} & \num{6.893}\\
    DSVI & Full rank   & \num{-13.97 \pm 0.04} & \num{4.024} & \num{5.968}\\
    & Chevron $k = 10$ & \num{-14.56 \pm 0.04}  & \num{4.007} & \num{6.265}\\
    & Chevron $k = 5$  & \num{-15.02 \pm 0.04}  & \num{4.343} & \num{6.778}\\
    & Chevron $k = 2$  & \num{-16.23 \pm 0.04}  & \num{4.768} & \num{7.516}\\
    & Mean field       & \num{-16.66 \pm 0.04}  & \num{5.353} & \num{9.297}\\
    \bottomrule
  \end{tabular}
  \caption{Reference and estimated hyperparameters, and simple MC estimate of the ELBO, for the one-dimensional nonlinear-diffusion problem}
  \label{tab:richards-theta}
\end{table}

\section{Conclusions and discussion}
\label{sec:conclusions}

We have presented two approximate empirical Bayesian methods, Laplace-EM and DSVI-EB, for estimating unknown parameters and constitutive relations in PDE models.
Compared to other methods for approximate Bayesian inference, the proposed methods do not require third-order derivatives of the physics model, do not involve computing moments of non-Gaussian likelihoods, and are applicable to non-factorizing likelihoods.
Furthermore, the calculation of the batch estimate of the ELBO and its gradients employed in the DSVI-EB method is trivially parallelizable, leading to savings in computational time.
The numerical experiments presented show that both methods accurately approximate the posterior density and the hyperparameters of the GP prior.
In particular, we find that the Laplace-EM method is more accurately approximate the posterior density, at the cost of computing Hessians of the physics model, which increase the computational cost of each EM cycle.
The DSVI-EB method, on the other hand, is less accurate but does not require Hessians.
Consistent with the literature, we find that the accuracy of the DSVI-EB method at approximating the posterior decreases with increasing sparsity of the covariance factor parameterization employed.

For a very large number of degrees of freedom of the discretization of the unknown functions, the computational cost of the proposed methods is dominated by the associated cubic complexity.
Future work will aim to address the challenge of cubic complexity by employing sparse GP inference.

\appendix

\section{Gaussian backpropagation rules}
\label{sec:gaussian-backprop}

In this section we present the derivation of the gradients~\eqref{eq:elbo-z-est-grad-muq}--\eqref{eq:elbo-z-est-grad-theta}.

For the gradient with respect to the variational mean, \eqref{eq:elbo-z-est-grad-muq}, we have by the chain rule, in index notation,
\begin{equation*}
  \frac{\partial}{\partial \mu_{q, i}} \log p(\mathcal{D}_s \mid \mathbf{y}) = \frac{\partial y_j}{\partial \mu_{q, i}} \frac{\partial}{\partial y_j} \log p(\mathcal{D}_s \mid \mathbf{y}) = \delta_{ji} \frac{\partial}{\partial y_j} \log p(\mathcal{D}_s \mid \mathbf{y}),
\end{equation*}%

where we have used $\partial y_j / \partial \mu_{q, i} = \delta_{ji}$.
It follows that $\nabla_{\bm{\mu}_q} \log p(\mathcal{D}_s \mid \mathbf{y}) = \nabla_{\mathbf{y}} \log p(\mathcal{D}_s \mid \mathbf{y})$.
Similarly, we have
\begin{equation*}
  \nabla_{\bm{\mu}_q} \frac{1}{2} \mathbf{y}^{\top} \mathbf{C}^{-1}_p \mathbf{y} = \nabla_{\mathbf{y}} \frac{1}{2} \mathbf{y}^{\top} \mathbf{C}^{-1}_p \mathbf{y} = \mathbf{C}^{-1}_p \mathbf{y},
\end{equation*}
and thus we recover~\eqref{eq:elbo-z-est-grad-muq}.

For the gradient with respect to the Cholesky factor $\mathbf{R}_q$, we note that $\partial y_k / \partial R_{q, ij} = \delta_{ki} z_j$.
By the chain rule, we have
\begin{equation*}
  \frac{\partial}{\partial R_{q, ij}} \log p(\mathcal{D}_s \mid \mathbf{y}) = \frac{\partial y_k}{\partial R_{q, ij}}\frac{\partial}{\partial y_k} \log p(\mathcal{D}_s \mid \mathbf{y}) = \frac{\partial}{\partial y_i} \log p(\mathcal{D}_s \mid \mathbf{y}) z_j,  
\end{equation*}
so that $\nabla_{\mathbf{R}_q} \log p(\mathcal{D}_s \mid \mathbf{y}) = \left [ \nabla_{\mathbf{y}} \log p(\mathcal{D}_s \mid \mathbf{y}) \right ] \mathbf{z}^{\top}$.
Similarly,
\begin{equation*}
  \nabla_{\mathbf{R}_q} \frac{1}{2} \mathbf{y}^{\top} \mathbf{C}^{-1}_p \mathbf{y} = \left [ \nabla_{\mathbf{y}} \frac{1}{2} \mathbf{y}^{\top} \mathbf{C}^{-1}_p \mathbf{y} \right ] \mathbf{z}^{\top} = \mathbf{C}^{-1}_p \mathbf{y} \mathbf{z}^{\top}.
\end{equation*}
Finally, we have $\nabla_{\mathbf{R}_q} \log \det \mathbf{R}_q = (\mathbf{R}^{-1}_q)^{\top}$, from which we recover~\eqref{eq:elbo-z-est-grad-Rq}.

The gradients with respect to the prior hyperparameters, \eqref{eq:elbo-z-est-grad-theta} can be derived from~\cite{rasmussen-2005-gaussian}, Eqs. (A.14) and (A.15).

For the gradient with respect to $\bm{\omega}_q$ of the mean-field parameterization $\mathbf{R}_q = \operatorname{diag}[\exp{\bm{\omega}_q}]$, we employ the relation
\begin{equation*}
  \frac{\partial R_{q, ij}}{\partial \omega_{q, k}} =
  \begin{cases}
    \exp \omega_{q, k} & \text{for } k = i = j,\\
    0 & \text{otherwise}.
  \end{cases}
\end{equation*}
By the chain rule, we have
\begin{equation*}
  \frac{\partial}{\partial \omega_{q, k}} \log p(\mathcal{D}_s \mid \mathbf{y}) = \frac{\partial R_{q, ij}}{\partial \omega_{q, k}} \frac{\partial}{\partial R_{q, ij}} \log p(\mathcal{D}_s \mid \mathbf{y}) = \frac{\partial}{\partial y_k} \log p(\mathcal{D}_s \mid \mathbf{y}) z_k \exp \omega_{q, k},
\end{equation*}
summation over $k$ not implied.
It follows that $\nabla_{\bm{\omega}_q} \log p(\mathcal{D}_s \mid \mathbf{y}) = \left [ \nabla_{\mathbf{y}} \log p(\mathcal{D}_s \mid \mathbf{y}) \right ] \circ \mathbf{z} \circ \exp{\bm{\omega}_q}$.
Similarly,
\begin{equation*}
  \frac{\partial}{\partial \omega_{q, k}} \frac{1}{2} \mathbf{y}^{\top} \mathbf{C}^{-1}_p \mathbf{y} = \frac{\partial R_{q, ij}}{\partial \omega_{q, k}} \left ( \mathbf{C}^{-1}_p \right )_{im} y_m z_j = \left ( \mathbf{C}^{-1}_p \right )_{km} y_m z_k \exp \omega_{q, k},
\end{equation*}
summation over $k$ not implied.
Finally,
\begin{equation*}
  \nabla_{\bm{\omega}_q} \log \det \mathbf{R}_q = \nabla_{\bm{\omega}_q} \log \prod_k \exp{\omega_{q, k}} = \nabla_{\bm{\omega}_q} \sum_k \log \exp{\omega_{q, k}} = \mathbf{I}_N,
\end{equation*}
from which we recover~\eqref{eq:elbo-z-est-grad-omegaq}.

\section{Stochastic gradient ascent with adaptive step-size sequence}
\label{sec:as3}

Here we reproduce for completeness the stochastic gradient ascent algorithm with adaptive step-size sequence proposed in~\cite{kucukelbir-2017-automatic}.
The presentation is expanded to the empirical Bayes context for the update of prior hyperparameters.
At each iteration, the variational parameters and prior hyperparameters are updated using the rules
\begin{align}
  \label{eq:as3-phi-update}
  \bm{\phi}^{(j + 1)} &= \bm{\phi}^{(j)} + \bm{\rho}^{(j)}_{\bm{\phi}} \circ \nabla_{\bm{\phi}} f^{(j)}_n,\\
  \label{eq:as3-theta-update}
  \bm{\theta}^{(j + 1)} &= \bm{\theta}^{(j)} + \bm{\rho}^{(j)}_{\bm{\theta}} \circ \nabla_{\bm{\theta}} f^{(j)}_n,
\end{align}
where $f^{(j)}_n \equiv f_n(\bm{\phi}^{(j)}, \bm{\theta}^{(j)})$, and the vectors of step-sizes $\bm{\rho}^{(j)}_{\bm{\phi}}$ and $\bm{\rho}^{(j)}_{\bm{\theta}}$ are given by
\begin{equation}
  \label{eq:as3-rho}
  \bm{\rho}^{(j)}_{\bm{\phi} \backslash \bm{\theta}} = \eta (j + 1)^{-1 / 2 + \epsilon} \left (\tau + \sqrt{\mathbf{s}^{(j)}_{\bm{\phi} \backslash \bm{\theta}}} \right ),
\end{equation}
and the sequence
\begin{equation}
  \label{eq:as3-s}
  \mathbf{s}^{(j)}_{\bm{\phi} \backslash \bm{\theta}} = \alpha \left (\nabla_{\bm{\phi} \backslash \bm{\theta}} f^{(j)}_n \right )^2 + (1 - \alpha) \mathbf{s}^{(j)}_{\bm{\phi} \backslash \bm{\theta}} \text{ for } j > 0, \quad s^{(0)}_{\bm{\phi} \backslash \bm{\theta}} = \left (\nabla_{\bm{\phi} \backslash \bm{\theta}} f^{(0)}_n \right )^2,
\end{equation}
where $\sqrt{\cdot}$ and $(\cdot)^2$ are understood as element-wise.
The parameters $\tau$, $\alpha$, and $\epsilon$ are set to $1.0$, $0.1$, and $\num{1e-16}$, respectively, while the parameter $\eta > 0$ is chosen on a case-by-case basis.

\section{Discrete adjoint method for Darcy flow}
\label{sec:da}

In this section we describe the computation of the gradient and Hessian of the log-likelihood, $\nabla_{\mathbf{y}} \log p(\mathcal{D}_s \mid \mathbf{y})$ via the discrete adjoint method~\cite{giles-2003-algorithm,ghate-2007-efficient}.
For this purpose we introduce the function
\begin{equation}
  \label{eq:da-h}
  h(\mathbf{u}, \mathbf{y}) = -\frac{1}{2 \sigma^2_{us}} \| \mathbf{u}_s - \mathbf{H}_u \mathbf{u} \|^2_2 - \frac{1}{2 \sigma^2_{ys}} \| \mathbf{y}_s - \mathbf{H}_y \mathbf{y} \|^2_2,
\end{equation}
so that $\nabla_{\mathbf{y}} \log p(\mathcal{D}_s \mid \mathbf{y}) = \nabla h(u(\mathbf{y}), \mathbf{y})$ by virtue of~\eqref{eq:likelihood} (as the constant in~\eqref{eq:likelihood} is independent of $\mathbf{y}$).
In the following we will employ the following notation: Let $a$ be a scalar function, $\mathbf{b}$ and $\mathbf{c}$ be vector functions, and $\gamma$ be a scalar variable; then, $\partial a / \partial \mathbf{b}$ denotes the row vector with entries $\partial a / \partial b_i$, $\partial \mathbf{b} / \partial \gamma$ denotes the column vector with entries $\partial b_i / \partial \gamma$, and $\partial \mathbf{b} / \partial \mathbf{c}$ be the matrix with $ij$th entry $\partial b_i / \partial c_j$.

Differentiation $h(\mathbf{u}, \mathbf{y})$ with respect to $y_i$ gives
\begin{equation}
  \label{eq:da-h-total}
  \frac{\mathrm{d} h}{\mathrm{d} y_j} = \frac{\partial h}{\partial y_j} + \frac{\partial h}{\partial \mathbf{u}} \frac{\partial \mathbf{u}}{\partial y_j}.
\end{equation}
Similarly, differentiating the physics constraint $\mathbf{L}(\mathbf{u}, \mathbf{y}) = 0$ with respect to $\mathbf{y}$ gives
\begin{equation}
  \label{eq:da-L-total}
  \frac{\partial \mathbf{L}}{\partial y_j} + \frac{\partial \mathbf{L}}{\partial \mathbf{u}} \frac{\partial \mathbf{u}}{\partial y_j} = 0
\end{equation}
which implies $\partial \mathbf{u} / \partial y_j = -(\partial \mathbf{L} / \partial \mathbf{u})^{-1} (\partial \mathbf{L} / \partial y_j)$.
Substituting this relation into~\eqref{eq:da-h-total} gives the following expression for the $j$th component of the gradient:
\begin{equation}
  \label{eq:da-h-adj-grad}
  \frac{\mathrm{d} h}{\mathrm{d} y_j} = \frac{\partial h}{\partial y_i} + \bm{\lambda}^{\top} \frac{\partial \mathbf{L}}{\partial y_j}
\end{equation}
where the adjoint variables $\bm{\lambda}$ satisfies the adjoint equation
\begin{equation}
  \label{eq:da-adj}
  \left (\frac{\partial \mathbf{L}}{\partial \mathbf{u}} \right )^{\top} \bm{\lambda} + \left ( \frac{\partial h}{\partial \bm{u}} \right )^{\top} = 0.
\end{equation}
It can be seen that computing the gradient $\nabla h(u(\mathbf{y}), \mathbf{y})$ requires a single linear backward sensitivity problem, \eqref{eq:da-adj}, of size $M \times M$.

For the Hessian, we differentiate~\eqref{eq:da-h-total} with respect to $y_i$, obtaining
\begin{equation}
  \label{eq:da-h-total-2}
  \frac{\mathrm{d}^2 h}{\mathrm{d} y_i \mathrm{d} y_j} = \frac{\partial h}{\partial \mathbf{u}} \frac{\partial^2 \mathbf{u}}{\partial y_i \partial y_j} + D^2_{i, j} h,
\end{equation}
where $D^2_{i, j} h$ is given by
\begin{equation}
  \label{eq:da-D2}
  \frac{\partial^2 h}{\partial y_i \partial y_j} + \frac{\partial^2 h}{\partial y_i \partial \mathbf{u}} \frac{\partial \mathbf{u}}{\partial y_j} + \frac{\partial^2 h}{\partial y_j \partial \mathbf{u}} \frac{\partial \mathbf{u}}{\partial y_i} + \frac{\partial^2 h}{\partial \mathbf{u}^2} \left ( \frac{\partial \mathbf{u}}{\partial y_i} \otimes \frac{\partial \mathbf{u}}{\partial y_j} \right ),
\end{equation}
and $\partial^2 h / \partial \mathbf{u}^2$ denotes the Hessian of $h$ with respect to $\mathbf{u}$, i.e. the matrix with $ij$th entry $\partial^2 h / \partial u_i \partial u_j$.
Similarly, differentiating~\eqref{eq:da-L-total} with respect to $y_i$ gives
\begin{equation}
  \label{eq:da-L-total-2}
  \frac{\partial \mathcal{L}}{\partial \mathbf{u}} \frac{\partial^2 \mathbf{u}}{\partial y_i \partial y_j} + D^2_{i, j} \mathbf{L} = 0,
\end{equation}
where $D^2_{i, j} \mathbf{L}$ is given element-wise in a manner similar to~\eqref{eq:da-D2}.
\eqref{eq:da-L-total-2} implies $\partial^2 \mathbf{u} / \partial y_i \partial y_j = - (\partial \mathbf{L} / \partial \mathbf{u})^{-1} D^2_{i, j} \mathbf{L}$.
Substituting into~\eqref{eq:da-h-total-2} gives the following expression for the $ij$th component of the Hessian:
\begin{equation}
  \label{eq:da-adj-Hessian}
  \frac{\mathrm{d}^2 h}{\mathrm{d} y_i \mathrm{d} y_j} = \bm{\lambda}^{\top} D^2_{i, j} \mathbf{L} + D^2_{i, j} h.
\end{equation}
Computing the Hessian therefore requires the solution of $N$ linear forward sensitivity problems, \eqref{eq:da-L-total}, for each $\partial \mathbf{u} / \partial y_i$, and one backward sensitivity solution for the adjoint variables, each problem of size $M \times M$.

\section*{Acknowledgments}

This work was supported by the Applied Mathematics Program within the U.S. Department of Energy Office of Advanced Scientific Computing Research.
Pacific Northwest National Laboratory is operated by Battelle for the DOE under Contract DE-AC05-76RL01830.

\bibliographystyle{elsarticle-num}

\begin{thebibliography}{10}
\expandafter\ifx\csname url\endcsname\relax
  \def\url#1{\texttt{#1}}\fi
\expandafter\ifx\csname urlprefix\endcsname\relax\def\urlprefix{URL }\fi
\expandafter\ifx\csname href\endcsname\relax
  \def\href#1#2{#2} \def\path#1{#1}\fi

\bibitem{stuart_2010}
A.~M. Stuart, Inverse problems: A bayesian perspective, Acta Numerica 19 (2010)
  451–559.
\newblock \href {http://dx.doi.org/10.1017/S0962492910000061}
  {\path{doi:10.1017/S0962492910000061}}.

\bibitem{hanke-1997-regularizing}
M.~Hanke, \href{http://stacks.iop.org/0266-5611/13/i=1/a=007}{A regularizing
  levenberg-marquardt scheme, with applications to inverse groundwater
  filtration problems}, Inverse Problems 13~(1) (1997) 79.
\newline\urlprefix\url{http://stacks.iop.org/0266-5611/13/i=1/a=007}

\bibitem{barajassolano-2014-linear}
D.~A. Barajas-Solano, B.~E. Wohlberg, V.~V. Vesselinov, D.~M. Tartakovsky,
  Linear functional minimization for inverse modeling, Water Resour. Res. 51
  (2014) 4516--4531.
\newblock \href {http://dx.doi.org/10.1002/2014WR016179}
  {\path{doi:10.1002/2014WR016179}}.

\bibitem{evensen-2006-data}
G.~Evensen, Data Assimilation: The Ensemble Kalman Filter, Springer-Verlag,
  Berlin, Heidelberg, 2006.

\bibitem{salimans-2015-markov}
T.~Salimans, D.~Kingma, M.~Welling,
  \href{http://proceedings.mlr.press/v37/salimans15.html}{Markov chain monte
  carlo and variational inference: Bridging the gap}, in: F.~Bach, D.~Blei
  (Eds.), Proceedings of the 32nd International Conference on Machine Learning,
  Vol.~37 of Proceedings of Machine Learning Research, PMLR, Lille, France,
  2015, pp. 1218--1226.
\newline\urlprefix\url{http://proceedings.mlr.press/v37/salimans15.html}

\bibitem{goodman-2010-ensemble}
J.~Goodman, J.~Weare, \href{https://doi.org/10.2140/camcos.2010.5.65}{Ensemble
  samplers with affine invariance}, Commun. Appl. Math. Comput. Sci. 5~(1)
  (2010) 65--80.
\newblock \href {http://dx.doi.org/10.2140/camcos.2010.5.65}
  {\path{doi:10.2140/camcos.2010.5.65}}.
\newline\urlprefix\url{https://doi.org/10.2140/camcos.2010.5.65}

\bibitem{neiswanger-2013-asymptotically}
W.~{Neiswanger}, C.~{Wang}, E.~{Xing}, {Asymptotically Exact, Embarrassingly
  Parallel MCMC}, ArXiv e-prints\href {http://arxiv.org/abs/1311.4780}
  {\path{arXiv:1311.4780}}.

\bibitem{hoffman-2014-nuts}
M.~D. Hoffman, A.~Gelman, \href{http://jmlr.org/papers/v15/hoffman14a.html}{The
  no-u-turn sampler: Adaptively setting path lengths in hamiltonian monte
  carlo}, Journal of Machine Learning Research 15 (2014) 1593--1623.
\newline\urlprefix\url{http://jmlr.org/papers/v15/hoffman14a.html}

\bibitem{rasmussen-2005-gaussian}
C.~E. Rasmussen, C.~K.~I. Williams, Gaussian Processes for Machine Learning
  (Adaptive Computation and Machine Learning), The MIT Press, 2005.

\bibitem{raissi-2018-numerical}
M.~Raissi, P.~Perdikaris, G.~E. Karniadakis,
  \href{https://doi.org/10.1137/17M1120762}{Numerical gaussian processes for
  time-dependent and nonlinear partial differential equations}, SIAM Journal on
  Scientific Computing 40~(1) (2018) A172--A198.
\newblock \href {http://arxiv.org/abs/https://doi.org/10.1137/17M1120762}
  {\path{arXiv:https://doi.org/10.1137/17M1120762}}, \href
  {http://dx.doi.org/10.1137/17M1120762} {\path{doi:10.1137/17M1120762}}.
\newline\urlprefix\url{https://doi.org/10.1137/17M1120762}

\bibitem{raissi-2017-machine}
M.~Raissi, P.~Perdikaris, G.~E. Karniadakis,
  \href{http://www.sciencedirect.com/science/article/pii/S0021999117305582}{Machine
  learning of linear differential equations using gaussian processes}, Journal
  of Computational Physics 348 (2017) 683 -- 693.
\newblock \href {http://dx.doi.org/https://doi.org/10.1016/j.jcp.2017.07.050}
  {\path{doi:https://doi.org/10.1016/j.jcp.2017.07.050}}.
\newline\urlprefix\url{http://www.sciencedirect.com/science/article/pii/S0021999117305582}

\bibitem{bishop-2006-pattern}
C.~M. Bishop, Pattern Recognition and Machine Learning (Information Science and
  Statistics), Springer-Verlag, Berlin, Heidelberg, 2006.

\bibitem{lawrence-2007-modelling}
N.~D. Lawrence, G.~Sanguinetti, M.~Rattray,
  \href{http://papers.nips.cc/paper/3119-modelling-transcriptional-regulation-using-gaussian-processes.pdf}{Modelling
  transcriptional regulation using gaussian processes}, in: B.~Sch\"{o}lkopf,
  J.~C. Platt, T.~Hoffman (Eds.), Advances in Neural Information Processing
  Systems 19, MIT Press, 2007, pp. 785--792.
\newline\urlprefix\url{http://papers.nips.cc/paper/3119-modelling-transcriptional-regulation-using-gaussian-processes.pdf}

\bibitem{neal-1998-view}
R.~M. Neal, G.~E. Hinton,
  \href{https://doi.org/10.1007/978-94-011-5014-9\_12}{A View of the {EM}
  Algorithm that Justifies Incremental, Sparse, and other Variants}, Springer
  Netherlands, Dordrecht, 1998, pp. 355--368.
\newblock \href {http://dx.doi.org/10.1007/978-94-011-5014-9_12}
  {\path{doi:10.1007/978-94-011-5014-9_12}}.
\newline\urlprefix\url{https://doi.org/10.1007/978-94-011-5014-9\_12}

\bibitem{ranganath-2014-black}
R.~{Ranganath}, S.~{Gerrish}, D.~{Blei}, Black box variational inference, in:
  S.~Kaski, J.~Corander (Eds.), Proceedings of the Seventeenth International
  Conference on Artificial Intelligence and Statistics, Vol.~33 of Proceedings
  of Machine Learning Research, PMLR, Reykjavik, Iceland, 2014, pp. 814--822.

\bibitem{titsias-2014-doubly}
M.~Titsias, M.~{L\'{a}zaro-Gredilla}, Doubly stochastic variational bayes for
  non-conjugate inference, in: E.~P. {Xing}, T.~{Jebara} (Eds.), Proceedings of
  the 31st International Conference on Machine Learning, Vol.~32 of Proceedings
  of Machine Learning Research, PMLR, Bejing, China, 2014, pp. 1971--1979.

\bibitem{minka-2001-expectation}
T.~P. Minka,
  \href{http://dl.acm.org/citation.cfm?id=2074022.2074067}{Expectation
  propagation for approximate bayesian inference}, in: Proceedings of the
  Seventeenth Conference on Uncertainty in Artificial Intelligence, UAI'01,
  Morgan Kaufmann Publishers Inc., San Francisco, CA, USA, 2001, pp. 362--369.
\newline\urlprefix\url{http://dl.acm.org/citation.cfm?id=2074022.2074067}

\bibitem{tsilifis-2016-computationally}
P.~Tsilifis, I.~Bilionis, I.~Katsounaros, N.~Zabaras, Computationally efficient
  variational approximations for bayesian inverse problems, Journal of
  Verification, Validation and Uncertainty Quantification 1~(3) (2016) 031004.

\bibitem{jin-2010-hierarchical}
B.~Jin, J.~Zou,
  \href{http://www.sciencedirect.com/science/article/pii/S0021999110003311}{Hierarchical
  bayesian inference for ill-posed problems via variational method}, Journal of
  Computational Physics 229~(19) (2010) 7317 -- 7343.
\newblock \href {http://dx.doi.org/https://doi.org/10.1016/j.jcp.2010.06.016}
  {\path{doi:https://doi.org/10.1016/j.jcp.2010.06.016}}.
\newline\urlprefix\url{http://www.sciencedirect.com/science/article/pii/S0021999110003311}

\bibitem{franck-2016-sparse}
I.~M. Franck, P.~Koutsourelakis,
  \href{http://www.sciencedirect.com/science/article/pii/S0045782515003345}{Sparse
  variational bayesian approximations for nonlinear inverse problems:
  Applications in nonlinear elastography}, Computer Methods in Applied
  Mechanics and Engineering 299 (2016) 215 -- 244.
\newblock \href {http://dx.doi.org/https://doi.org/10.1016/j.cma.2015.10.015}
  {\path{doi:https://doi.org/10.1016/j.cma.2015.10.015}}.
\newline\urlprefix\url{http://www.sciencedirect.com/science/article/pii/S0045782515003345}

\bibitem{guha-2015-variational}
N.~Guha, X.~Wu, Y.~Efendiev, B.~Jin, B.~K. Mallick,
  \href{http://www.sciencedirect.com/science/article/pii/S002199911500515X}{A
  variational bayesian approach for inverse problems with skew-t error
  distributions}, Journal of Computational Physics 301 (2015) 377 -- 393.
\newblock \href {http://dx.doi.org/https://doi.org/10.1016/j.jcp.2015.07.062}
  {\path{doi:https://doi.org/10.1016/j.jcp.2015.07.062}}.
\newline\urlprefix\url{http://www.sciencedirect.com/science/article/pii/S002199911500515X}

\bibitem{yang-2017-bayesian}
K.~Yang, N.~Guha, Y.~Efendiev, B.~K. Mallick,
  \href{http://www.sciencedirect.com/science/article/pii/S0021999117303054}{Bayesian
  and variational bayesian approaches for flows in heterogeneous random media},
  Journal of Computational Physics 345 (2017) 275 -- 293.
\newblock \href {http://dx.doi.org/https://doi.org/10.1016/j.jcp.2017.04.034}
  {\path{doi:https://doi.org/10.1016/j.jcp.2017.04.034}}.
\newline\urlprefix\url{http://www.sciencedirect.com/science/article/pii/S0021999117303054}

\bibitem{blei-2017-variational}
D.~M. Blei, A.~Kucukelbir, J.~D. McAuliffe,
  \href{https://doi.org/10.1080/01621459.2017.1285773}{Variational inference: A
  review for statisticians}, Journal of the American Statistical Association
  112~(518) (2017) 859--877.
\newblock \href
  {http://arxiv.org/abs/https://doi.org/10.1080/01621459.2017.1285773}
  {\path{arXiv:https://doi.org/10.1080/01621459.2017.1285773}}, \href
  {http://dx.doi.org/10.1080/01621459.2017.1285773}
  {\path{doi:10.1080/01621459.2017.1285773}}.
\newline\urlprefix\url{https://doi.org/10.1080/01621459.2017.1285773}

\bibitem{kucukelbir-2017-automatic}
A.~{Kucukelbir}, D.~{Tran}, R.~{Ranganath}, A.~{Gelman}, D.~M. {Blei},
  Automatic differentiation variational inference, Journal of Machine Learning
  Research 18~(14) (2017) 1--45.

\bibitem{kingma-2013-auto}
D.~P. {Kingma}, M.~{Welling}, {Auto-Encoding Variational Bayes}, ArXiv
  e-prints\href {http://arxiv.org/abs/1312.6114} {\path{arXiv:1312.6114}}.

\bibitem{rezende-2014-stochastic}
D.~J. {Rezende}, S.~{Mohamed}, D.~{Wierstra}, {Stochastic Backpropagation and
  Approximate Inference in Deep Generative Models}, ArXiv e-prints\href
  {http://arxiv.org/abs/1401.4082} {\path{arXiv:1401.4082}}.

\bibitem{challis-2013-gaussian}
E.~Challis, D.~Barber,
  \href{http://jmlr.org/papers/v14/challis13a.html}{Gaussian kullback-leibler
  approximate inference}, Journal of Machine Learning Research 14 (2013)
  2239--2286.
\newline\urlprefix\url{http://jmlr.org/papers/v14/challis13a.html}

\bibitem{willliams-1992-simple}
R.~J. {Williams}, Simple Statistical Gradient-Following Algorithms for
  Connectionist Reinforcement Learning, Springer US, Boston, MA, 1992, pp.
  5--32.

\bibitem{friston-2007-variational}
K.~Friston, J.~Mattout, N.~Trujillo-Barreto, J.~Ashburner, W.~Penny,
  \href{http://www.sciencedirect.com/science/article/pii/S1053811906008822}{Variational
  free energy and the laplace approximation}, NeuroImage 34~(1) (2007) 220 --
  234.
\newblock \href
  {http://dx.doi.org/https://doi.org/10.1016/j.neuroimage.2006.08.035}
  {\path{doi:https://doi.org/10.1016/j.neuroimage.2006.08.035}}.
\newline\urlprefix\url{http://www.sciencedirect.com/science/article/pii/S1053811906008822}

\bibitem{giles-2003-algorithm}
M.~B. {Giles}, M.~C. {Duta}, J.-D. {M\"{u}ller}, N.~A. {Pierce}, Algorithm
  developments for discrete adjoint methods, AIAA Journal 41~(2) (2003)
  198--205.

\bibitem{ghate-2007-efficient}
D.~Ghate, M.~Giles, Efficient hessian calculation using automatic
  differentiation, in: 25th AIAA Applied Aerodynamics Conference, 2007, p.
  4059.
\newblock \href {http://dx.doi.org/10.2514/6.2007-4059}
  {\path{doi:10.2514/6.2007-4059}}.

\end{thebibliography}

\end{document}